\documentclass[twocolumn,amsmath,amssymb,floatfix,superscriptaddress,pra,showpacs,letterpaper]{revtex4}

\usepackage{amsmath,amssymb,natbib,bm,graphicx,url,times}
\usepackage{color,multirow}
\usepackage[utf8]{inputenc}

\newcounter{fig}

\newcommand{\RR}{\mathbb{R}}
\newcommand{\NN}{\mathbb{N}}
\newcommand{\ZZ}{\mathbb{Z}}

\newcommand{\vc}[1]{\mathbf{#1}}

\newcommand{\abs}[1]{\left|#1\right|}
\newcommand{\bra}[1]{\left\langle \, #1 \,\right|}
\newcommand{\ket}[1]{\left|\, #1 \, \right\rangle}
\newcommand{\bket}[2]{\left\langle \, #1 \,|\, #2 \, \right\rangle}
\newcommand{\boket}[3]{\langle\, #1 \,|\, #2 \,|\, #3 \,\rangle}

\newcommand{\be}{\begin{equation}}
\newcommand{\ee}{\end{equation}}

\newcommand{\osum}{\circlearrowleft\hspace{-1.2em}\sum}

\begin{document}

\title{Time-reversal symmetry breaking in circuit-QED based photon lattices}
\author{Jens Koch}
\altaffiliation[Permanent address:]{Department of Physics and Astronomy, Northwestern University, Evanston, Illinois 60208, USA}
\affiliation{Departments of Physics and Applied Physics, Yale University, PO Box 208120, New Haven, CT 06520, USA}
\author{Andrew A.\ Houck}
\affiliation{Department of Electrical Engineering, Princeton University, Princeton, New Jersey 08544, USA}
\author{Karyn Le Hur}
\affiliation{Departments of Physics and Applied Physics, Yale University, PO Box 208120, New Haven, CT 06520, USA}
\author{S.\ M.\ Girvin}
\affiliation{Departments of Physics and Applied Physics, Yale University, PO Box 208120, New Haven, CT 06520, USA}

\begin{abstract}
Breaking time-reversal symmetry is a prerequisite for accessing certain interesting many-body states such as fractional quantum Hall states.
For polaritons, charge neutrality prevents magnetic fields from providing a direct symmetry breaking mechanism and similar to the situation in ultracold atomic gases, an effective magnetic field has to be synthesized. We show that in the circuit QED architecture, this can be achieved by inserting simple superconducting circuits into the resonator junctions. In the presence of such coupling elements, constant parallel magnetic and electric fields suffice to break time-reversal symmetry. We support these theoretical predictions with numerical simulations for realistic sample parameters, specify general conditions under which time-reversal is broken, and discuss the application to chiral Fock state transfer, an on-chip circulator, and tunable band structure for the Kagome lattice.
\end{abstract}

\date{June 3, 2010} 
\maketitle

\section{Introduction}
\begin{figure}
	\centering
		\includegraphics[width=1.0\columnwidth]{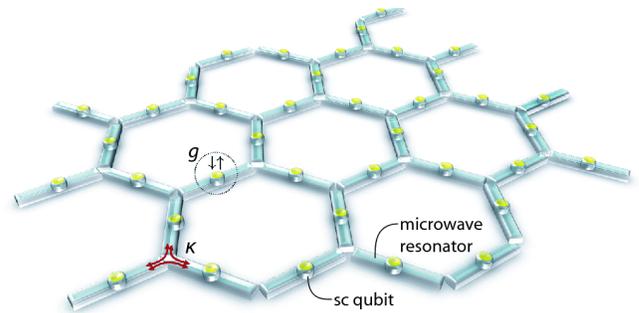}
	\caption{(Color online) The Jaynes-Cummings lattice as an example of a photon lattice. Its circuit QED realization would consist of superconducting resonators (e.g., coplanar waveguides, schematically shown as rectangular boxes), each of which would be coupled to a superconducting qubit (symbolized as dots centered in the resonators). Microwave photons would hop between nearest-neighbor resonators, with the coupling strength $\kappa$ set by the mutual capacitance between resonator ends. Interaction between the photons and the superconducting qubits with strength $g$ would induce an effective photon-photon interaction.\label{fig:JClattice}}
\end{figure}

Since the first pioneering papers in 2006 \cite{greentree_quantum_2006,hartmann_strongly_2006,angelakis_photon-blockade-induced_2007}, theoretical interest in the many body-physics of interacting photons or polaritons in lattices has flourished. Such photon lattices, see Fig.\ \ref{fig:JClattice} for an example,  are perceived as an interesting venue for quantum simulation \cite{buluta_quantum_2009} and for studying strongly correlated systems composed of polaritons \cite{hartmann_quantum_2008,le_hur_quantum_2009,tomadin_many-body_2010}. Hopes are that, once realized in experiments, such systems could complement the achievements in  research with ultracold atomic gases \cite{lewenstein_ultracold_2007,bloch_many-body_2008}, which are currently leading the charge.

 Much recent work has focused on the quantum phase transition between polaritonic Mott-insulating and superfluid states using various approaches  \cite{greentree_quantum_2006,hartmann_strongly_2006,angelakis_photon-blockade-induced_2007,rossini_mott-insulating_2007,rossini_photon_2008,na_strongly_2008-1,aichhorn_quantum_2008,zhao_insulator_2008,koch_superfluidmott-insulator_2009,schmidt_strong_2009-1,schmidt_excitations_2010-1}, and at this point there seems little doubt that the quantum phase transition is in the same universality class as its counterpart in the Bose-Hubbard model \cite{fisher_boson_1989,sachdev_quantum_2000,bruder_bose-hubbard_2005}. It is thus natural to ask, what physics beyond Bose-Hubbard might photon lattices have to offer?

 Recent work by several groups has highlighted the interesting implications of dissipation and external driving, and thus promoted the quantum phase transition to a nonequilibrium phase transition between different possible steady states \cite{tomadin_non-equilibrium_2009,carusotto_fermionized_2009,kiffner_dissipation-induced_2010,hartmann_polariton_2010}. A second route to physics beyond Bose-Hubbard, is to explore phases with broken time-reversal symmetry, of which fractional Quantum Hall phases are the most celebrated example \cite{tsui_two-dimensional_1982,laughlin_anomalous_1983-1}. 
 
To access such phases, a technique for breaking time-reversal symmetry is required. In contrast to electron gases, but similar to ultracold atomic gases \cite{jaksch_creation_2003,paredes_fractional_2003,srensen_fractional_2005,lin_synthetic_2009}, polariton systems face a challenge when trying to break time-reversal symmetry: due to the charge neutrality of polaritons, an external magnetic field cannot readily be used to achieve breaking of time-reversal, and instead an effective magnetic field has to be synthesized. A first proposal for cavity arrays with trapped three-level atoms and involving ac driving with specific phases has been published by Cho et al.\ \cite{cho_fractional_2008}. In addition, photonic edge states and analogs of the quantum Hall effect in photonic crystals have recently been investigated by Haldane and Raghu \cite{haldane_possible_2008,raghu_analogs_2008} and also probed experimentally \cite{zwang}.

In the present paper, we demonstrate that in the circuit QED architecture \cite{blais_cavity_2004,wallraff_strong_2004,schoelkopf_wiring_2008} breaking of time-reversal symmetry can be achieved by inserting simple superconducting circuits into resonator junctions and applying purely dc electric and magnetic fields. In our scheme, photons are transferred from resonator to resonator via virtual intermediate excitations of coupler circuits. We expect that the use of passive coupling elements and the absence of any ac fields pumping internal levels may avoid some of the challenges posed by dissipation. Our analysis shows that for broken particle-hole symmetry (caused by a dc electric field), polaritons can acquire an effective gauge charge and hence become susceptible to an external magnetic field so that time-reversal symmetry is broken. 
We emphasize that such passive coupling elements correspond to an important step towards substituting commercial microwave circulators  with on-chip circulators much smaller in size. This could pave the way for integrating circulators into larger arrays of resonators and could open interesting and new perspectives for correlated polariton systems. 

The remainder of the paper is organized as follows. In Section \ref{sec:general} we explain the generic consequences of integrating passive coupling elements into a resonator array and using them to break time-reversal symmetry. The passivity condition allows us to adiabatically eliminate the coupling elements and to  obtain an effective photonic tight-binding model with broken time-reversal symmetry. We emphasize the gauge-invariant phase sum (mimicking the contour integral of the magnetic vector potential in the continuous case) as a useful concept for determining whether time-reversal invariance holds. Applications of such coupling elements, including the prospect of an on-chip circulator conclude the section. 

Section \ref{sec:realization} then details our proposal for a physical realization of passive coupling elements in the circuit QED architecture. Specifically, we consider a system consisting of coplanar waveguide resonators which capacitively couple to small superconducting rings interrupted by three Josephson junctions (``Josephson rings"), which are inserted into the junctions between resonators. Using circuit quantization, we derive the Hamiltonian of this system and discuss the diagonalization of the Josephson rings. 

In Section \ref{sec:photonH}, we finally show how the adiabatic elimination of the ring degrees of freedom yields an effective photon Hamiltonian of the desired type. We discuss the general requirements for achieving time-reversal symmetry breaking in this scheme, and present results from numerical simulations which underline the proposal's feasibility with realistic device parameters. 

We end with conclusions and an outlook in Section \ref{sec:conclusions}. Some additional details of calculations and a self-contained summary of time-reversal symmetry in quantum mechanics are provided in several appendices. 

\begin{figure}
	\centering
		\includegraphics[width=0.5\columnwidth]{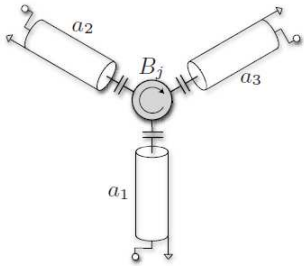}
	\caption{Basic scheme of a three-port coupling element, connected capacitively to three transmission-line resonators with annihilation operators $a_j$  for photons in the relevant mode of the resonators enumerated by $j=1,2,3$. \label{fig:circulator}}
\end{figure}

\section{Passive coupling elements for breaking time-reversal symmetry\label{sec:general}}
For the general discussion of breaking time-reversal symmetry by utilizing virtual excitations of a coupler circuit, we consider a junction
composed of three resonators \footnote{The restriction to a junction of three resonators is not essential, and can easily be generalized to larger resonator numbers.} coupled to a central ``circulator" system, see Fig.\ \ref{fig:circulator}, and described by a generic Hamiltonian of the form
\be\label{h1}
H=\sum_{j=1}^3 \omega_r a_j^\dag a_j + \lambda\sum_{j=1}^3(a_j+a_j^\dag)B_j +H_B.
\ee
 Here, $a_j$ and $a_j^\dag$ ($j=1,2,3$) are annihilation and creation operators for photons in the relevant mode of resonator $j$, with corresponding (angular) frequency $\omega_r$. (Note that throughout the paper we use units with $\hbar=1$.) The capacitive coupling between resonators and the degrees of freedom $B_j$ of the coupling element is described by the second term in Eq.\ \eqref{h1}. 

We  shall assume that the coupling element remains \emph{passive}, i.e.\ the coupler only transfers photons via intermediate virtual excitations and otherwise remains in its ground state at all times. Consequently, the coupler degrees of freedom can be integrated out (or, in other words, eliminated by a canonical transformation of Schrieffer-Wolff type \cite{schrieffer_relation_1966,cohen-tannoudji_atom-photon_1998}) so that one obtains an effective photon Hamiltonian $H_\text{eff}(a_j,a_j^\dag)$. The details of the effective Hamiltonian $H_\text{eff}$ generally depend on the specific realization of the passive coupling element, and we will go through the explicit derivation of $H_\text{eff}$ for the circuit QED realization we propose in Section \ref{sec:realization}. Here, we first explore the \emph{generic} properties of the effective photon Hamiltonian. 

We are interested in a passive coupling element that does not destroy the three-fold symmetry of the system. As a result,  there is a gauge in which $H_\text{eff}$ is invariant with respect to cyclic permutations of the indices $j=1,2,3$. Further, we assume that $H_\text{eff}$ allows for hopping of photons between resonators, but does not induce photon-photon interaction. (This assumption is realistic, as we show in Section \ref{sec:realization}.) As a result, $H_\text{eff}$ is anticipated to be a quadratic form of the annihilation and creation operators $a_j$, $a_j^\dag$. Explicitly, the Hamiltonian will take the form
\be\label{heff}
H_\text{eff}=\bigg[t(a_1a_3^\dag
+a_3a_2^\dag+a_2a_1^\dag)+\text{H.c.}\bigg] + \sum_{j=1}^3 \omega_r'a_j^\dag a_j ,
\ee
where $\omega_r'$ denotes the resonator frequency (possibly including a renormalization), and $t=\kappa\, e^{i\varphi}$ ($\kappa=\abs{t}\ge0$)  is the complex-valued hopping matrix element for photons \footnote{The fact that there is only one relevant phase $\varphi$ can readily be verified by employing a gauge transformation $a_j\to e^{i\varphi_j} a_j$.}.

When does the effective Hamiltonian \eqref{heff} describe the situation of broken time-reversal symmetry and when does time-reversal symmetry remain intact? Formally, time-reversal symmetry holds whenever the time-reversal operator $\Theta$ leaves the Hamiltonian invariant, i.e.\ $\Theta H \Theta^{-1}= H$ \footnote{Here, we are excluding the case of degenerate eigenstates of $H$, for which $\Theta$ can additionally induce a rotation within the degenerate subspace.}. As detailed in Appendix \ref{app:treversal}, for the present case this is true if there is a gauge transformation of the form
\be\label{gtrans1}
a_j\to e^{-i\varphi_j}a_j,
\ee
which makes the Hamiltonian real-valued when represented in the photon number basis. For the three-resonator junction, the existence of such a gauge transformation is checked as follows. According to Eqs.\ \eqref{heff} and \eqref{gtrans1}, an attempt to find a gauge transformation to make the Hamiltonian real-valued leads to the three equations
\begin{align}
\varphi + \varphi_1 - \varphi_3 = z_1 \pi,\nonumber\\\label{phi-equations}
\varphi + \varphi_2 - \varphi_1 = z_2 \pi,\\\nonumber
\varphi + \varphi_3 - \varphi_2 = z_3 \pi.
\end{align}
where $z_1,z_2,z_3\in \ZZ$ are arbitrary integers. These equations for the gauge phases $\varphi_1,\,\varphi_2$ and $\varphi_3$,  can only be solved (and hence time-reversal symmetry is intact) if the condition
\be\label{varphi}
3\varphi = z\pi \qquad (z\in\ZZ),
\ee
obtained by summing the three equations \eqref{phi-equations}, holds.  Thus, for the present case of a three-resonator junction we find: time-reversal symmetry is intact if and only if $\varphi\in\frac{\pi}{3}\ZZ$.

\begin{figure}
\centering
		\includegraphics[width=0.35\columnwidth]{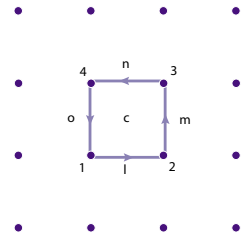}
	\caption{(Color online) Illustration of the gauge-invariant phase sum around a loop, 
	$\osum_{\mathcal{C}[ij]}\varphi_{ij}=\varphi_{12}+\varphi_{23}+\varphi_{34}+\varphi_{41}$,
	here for a particular plaquette $\mathcal{C}$ in a two-dimensional quadratic lattice.\label{fig:plaquette}}
\end{figure}

To extend this statement to general photon lattices with more resonators,
\be
H_\text{eff}=\sum_{i\not=j} t_{ij} a_i a_j^\dag + \sum_j \omega_r a_j^\dag a_j \qquad (t_{ji}=t_{ij}^*)
\ee
 it is important to identify the phase in Eq.\ \eqref{varphi} as a gauge-invariant quantity, which for discrete lattices plays a role analogous to the contour integral $\oint d\vc{s}\cdot\vc{A}$  of the vector potential $\vc{A}$ in the continuous case. [For simpler notation the prime in $\omega_r'$ has been dropped.] We write the gauge-invariant phase sum in the form
\be
\osum_{\mathcal{C}[ij]} \varphi_{ij} = \arg \prod_{\mathcal{C}[ij]} t_{ij},
\ee
where $\mathcal{C}$ specifies a closed path in the discrete lattice; see Fig.\ \ref{fig:plaquette} for an illustration. In these terms, the statement of Eq.\ \eqref{varphi} can be extended to larger systems where time-reversal symmetry can be shown to be intact if and only if the gauge-invariant phase sum is an integer multiple of $\pi$,
\be
\osum_{\mathcal{C}[ij]} \varphi_{ij}\in \pi\ZZ
\ee
for any closed lattice path $\mathcal{C}$.

To illustrate the implications of broken time-reversal symmetry, we discuss three examples: the clockwise or counterclockwise state transfer of a single photon Fock state between resonators, circulator behavior for signals propagating in semi-infinite transmission lines, and tunability of the Kagome tight-binding band structure. These examples are realizations of the simplest setting possible: resonators coupled via coupling elements and without any photon-photon interaction. The fascinating scenario of systems of \emph{interacting} photons with broken time-reversal symmetry is beyond the scope of this article and will be addressed in a future paper.

\subsection{Chiral transfer of photon Fock states}
We consider the 3-resonator junction depicted in Fig.\ \ref{fig:circulator} and described by the effective Hamiltonian $H_\text{eff}$, Eq. \eqref{heff}. $H_\text{eff}$ can be understood as a miniature tight-binding model with periodic boundary conditions.  The  eigenstates of $H_\text{eff}$ are generated by the creation operators
\be\label{ak}
A_k^\dag=\frac{1}{\sqrt{3}}\sum_{j=1}^3 e^{2\pi i k j / 3} a_j^\dag \qquad 
\ee
and have corresponding eigenenergies 
\be\label{omegak}
\Omega_k = \omega_r + 2\kappa \cos(2\pi k/3 +\varphi).
\ee
Here, $2\pi k/3$ ($k=-1,0,1$) are the allowed wave numbers in the first Brillouin zone. Recalling from Eq.\ \eqref{varphi} that time-reversal symmetry only holds as long as $\varphi\in\frac{\pi}{3}\ZZ$, it is not surprising that the simplest case of broken time-reversal symmetry (where the energy spectrum set by $\Omega_k$  becomes equidistant) is realized when $\varphi=\pm\pi/6$, i.e.\ halfway in between the time-reversal symmetric points $\varphi=0$ and $\pm\pi/3$.

\begin{figure}
	\centering
		\includegraphics[width=0.95\columnwidth]{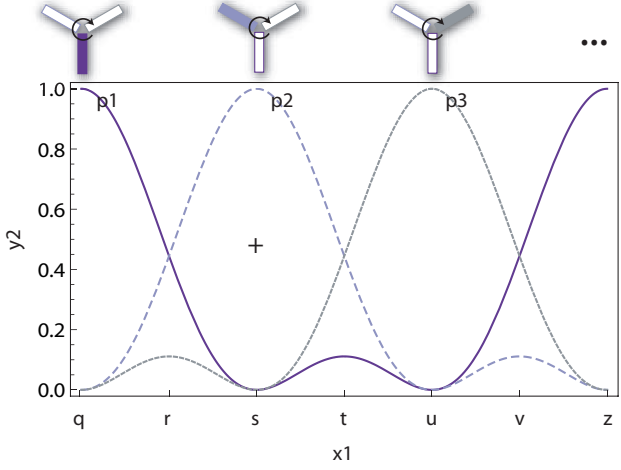}
	\caption{(Color online) Time evolution of a single-photon Fock state in the presence of a coupler with phase $\varphi=\pi/6$. The quantum state at the initial time $t=0$ is a Fock state with one photon in resonator $j=1$, and both resonators 2 and 3 in the vacuum state. The photon occupation probabilities $P_j$ are plotted as a function of time and show how the photon is transferred around the loop in a direction specified by the sign in $\varphi=\pm\pi/6$. The evolution is periodic with period $\tau=2\pi/\sqrt{3}\kappa$ and the initial state is transferred into Fock state of resonators 2 and 3 at times $t=\tau/3$ and $t=2\tau/3$, respectively.  \label{fig:photon-hop}}
\end{figure}

To understand the effect of broken time-reversal symmetry, let us consider the dynamics of the system inside the one-photon subspace. We initialize the system in a Fock state with a single photon inside one resonator, say resonator $j=1$, and follow its subsequent evolution in time.  The evolution  is obtained by
 solving the time-dependent Schr\"odinger equation with initial condition $\ket{\psi(t=0)}=a_{j=1}^\dag\ket{0}$. By using the inverse of the discrete Fourier transform  in Eq.\ \eqref{ak}, the evolution for $\varphi=\pm\pi/6$ is readily found to be
 \be\label{time-evolve}
 \ket{\psi(t)}= \frac{1}{\sqrt{3}}e^{i\omega_r t} \sum_{k=-1}^1 e^{ik\sqrt{3}\kappa t-2\pi i k/3}\ket{\psi_k},
\ee
where $\ket{\psi_k}\equiv A^\dag_k\ket{0}$ denotes the single-photon eigenstates of $H_\text{eff}$. The dynamics may be visualized  by plotting the probabilities 
\be
P_j(t)=\abs{\boket{0}{a_{j=1}}{\psi(t)}}^2
\ee
 for finding the photon in resonator $j$, see Fig.\ \ref{fig:photon-hop}. As expected from Eq.\ \eqref{time-evolve}, the dynamics is periodic with period $\tau = 2\pi/\sqrt{3}\kappa$. More importantly however, the breaking of time-reversal symmetry results in chirality: the photon is transferred from resonator to resonator either clockwise  or counter-clockwise depending on the sign of $\varphi=\pm\pi/6$.

\subsection{On-chip Circulator}
Circulators are lossless microwave elements with three (or more) ports, and have the crucial property that a signal entering port $j$ is fully transferred clockwise to port $j+1$ (or, alternatively counter-clockwise to port $j-1$) \cite{pozar_microwave_2004}. This behavior must involve breaking of time-reversal symmetry, which is typically accomplished by embedding magnetic material, e.g.\ ferrite, in the device. Commercial ferrite circulators are typically large ($\agt 1\,\text{cm}$) and their size would make it rather difficult to include large numbers in a photon lattice. It is thus interesting to explore the design of an \emph{on-chip circulator}, sufficiently small in size and easy to fabricate, such that it could be included in large numbers. In addition to being essential for breaking time-reversal symmetry in polariton lattices, such devices would find great practical application in the circuit QED architecture for quantum information processing.

Let us demonstrate that circulator behavior in the sense of microwave engineering can indeed be achieved with the model Hamiltonian $H_\text{eff}$, Eq.\ \eqref{heff}. The actual physical realization within the circuit QED architecture will be discussed in Section \ref{sec:realization}.   For simplicity, we consider a setting where microwave radiation is fed into the system by capacitively coupling semi-infinite transmission lines to the three resonators shown in Fig.\ \ref{fig:circulator}. The full system is then captured by the Hamiltonian 
\begin{align}\nonumber
H=&\omega_r\sum_{j=1}^3 a_j^\dag a_j  + \bigg[ \kappa e^{i\varphi}(a_1a_2^\dag+a_2a_3^\dag+a_3a_1^\dag)+\text{H.c.}\bigg]\\
&+\sum_{j=1}^3\sum_q \omega_q b_{jq}^\dag b_{jq} -i\sum_{j=1}^3\sum_q \left[f_qb_{jq}a_j^\dag-\text{H.c.}\right],
\end{align}
where $b_{jq}$ are the annihilation operators for the three transmission lines $j=1,2,3$, and $q$ is the mode index.

We divide the full Hamiltonian $H=H_\text{eff}+H_\text{tl}+H_\text{in}$ into the effective photon Hamiltonian previously discussed, the contribution from the semi-infinite transmission lines, and the interaction between them. Next, we employ the diagonalization of $H_\text{eff}$, see Eqs.\ \eqref{ak} and \eqref{omegak}, and rewrite the coupling Hamiltonian $H_\text{int}$ in terms of the eigenmodes $A_k$,
\be
H_\text{int} = -i\frac{1}{\sqrt{3}}\sum_q\sum_{j=1}^3\sum_{k=-1}^1 \bigg[ f_q
e^{-2\pi i j k/3}b_{jq} A_k^\dag -\text{H.c.}\bigg].
\ee
To calculate ingoing and outgoing fields, we use input-output theory \cite{walls_quantum_1995,clerk_introduction_2008}. As usual, formal integration of the Heisenberg equation of motion for $b_{jq}$,
\begin{align}
\dot{b}_{jq} =& - i\omega_q b_{jq} +\frac{1}{\sqrt{3}}f_q^* \sum_{k=-1}^1 e^{2\pi i j k /3}A_k,
\end{align}
yields solutions which can refer to either an initial state at time $t_i=t_0$ in the distant past, or to a final state at time $t_i=t_1$ in the distant future:
\begin{align}\nonumber
b_{jq}(t) =& e^{-i\omega_q(t-t_i)}b_{jq}(t_i) \\
           & +\frac{1}{\sqrt{3}} \int_{t_i}^t d\tau\, e^{-i\omega_q(t-\tau)}f_q^*\sum_{k=-1}^1 e^{2\pi i j k /3} A_k(\tau).\label{bjq}
\end{align}
Proceeding with standard input-output theory, we approximate the coupling matrix elements $f_q$ as constants within the relevant frequency range near $\Omega_k$, and employ the Markov approximation \cite{walls_quantum_1995}. 
We then plug Eq.\ \eqref{bjq} into the equation of motion for $A_k$,
\begin{align}
\dot{A}_k =& -i \Omega_k A_k -\frac{1}{\sqrt{3}}\sum_q \sum_{j=1}^3 f_q e^{-2\pi i j k/3} b_{jq}
\end{align}
 and identify
the input and output modes as
\begin{align}\nonumber
b_j^\text{in,out}(t) &= \frac{1}{\sqrt{2\pi\rho}}\sum_q e^{-i\omega_q(t-t_{0,1})}b_{jq}(t_{0,1}).
\end{align}
Here, $\rho$ is the transmission line density of states, and $\kappa'=2\pi\abs{f}^2\rho$ defines the effective photon decay rate. 
Applying the Markov approximation to the remaining time integral \cite{clerk_introduction_2008}, one obtains
\begin{align}\nonumber
\dot{A}_k =& -i\Omega_k A_k -\sqrt{\kappa'/3}\sum_{j=1}^3 e^{-2\pi i j k/3}b_j^\text{in}\\
&-\frac{\kappa'}{6}\sum_{k'=-1}^1 \sum_{j=1}^3 e^{-2\pi i j (k-k')/3}A_{k'}.\label{adot}
\end{align}
Analogous expressions for $\dot{A}_k$ (but with a crucial sign change in the last term) can be obtained when substituting the outgoing fields. By subtracting from Eq.\ \eqref{adot} the equations obtained when using the outgoing field in either port 1, 2, or 3, one can derive the following relation between ingoing and outgoing modes:
\be
b_j^\text{out} = b_j^\text{in} + \sqrt{\kappa/3}\sum_{k=-1}^1 e^{2\pi i j k/3}A_{k}.
\ee
\begin{figure}
	\centering
		\includegraphics[width=0.7\columnwidth]{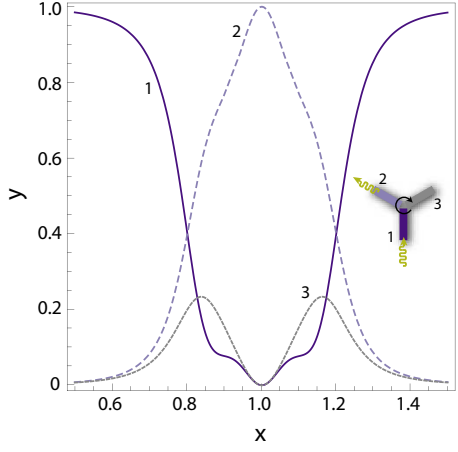}
	\caption{Circulator behavior. The plot shows the normalized outgoing power $\abs{b_{j}^\text{out}/b_1^\text{in}}^2$ for the three ports $j=1,2,3$ under coherent driving of port 1 with frequency $\omega_d$ when the phase $\varphi$ of the coupler element is adjusted to $\pi/6$. For drive frequencies close to the resonator frequency, the signal is transferred from port 1 to port 2. The bandwidth of this circulator behavior is set by the photon hopping rates ($\kappa=\kappa'/2=0.1\omega_r$).
	\label{fig:kappa}}
\end{figure}
Finally, we eliminate the dependence on the circulator modes by substituting the solutions to Eq.\ \eqref{adot}, which in frequency space can be expressed as
\be
A_k[\omega]=\frac{\sqrt{\kappa'/3}}{i(\omega-\Omega_k)-\kappa'/2}\sum_{j=1}^3 e^{-2\pi i j k/3}b_j^\text{in}[\omega],
\ee
 In total, one thus obtains the relation
\begin{align}\label{inout}
&b_j^\text{out}[\omega]=b_j^\text{in}[\omega]\\\nonumber
&\quad +\frac{\kappa'}{3}\sum_{k=-1}^1\sum_{j'=1}^3\frac{e^{2\pi i (j-j') k/3}}{i(\omega-\Omega_k)-\kappa'/2}  b_{j'}^\text{in}[\omega]
\end{align}
between the ingoing and outgoing fields. For coherent driving with frequency $\omega_d$, the ingoing and outgoing fields are characterized by c-numbers $\langle b_{j}^\text{in,out}[\omega_d]\rangle$ and the normalized outgoing power can be calculated from Eq.\ \eqref{inout}. Assuming a drive at only one of the input ports, say port 1, the normalized  outgoing power on port $j$ is $\abs{\langle b_{j}^\text{out}[\omega_d]\rangle/\langle b_{1}^\text{in}[\omega_d]\rangle}^2$. Note that results for driving on any other port can be obtained by cyclic permutation of the port indices)

As shown in Fig.\ \ref{fig:kappa}, the device shows clear circulator behavior when choosing $\varphi=\pi/6$ for the photon hopping phase. The circulator behavior is strongest when the drive frequency $\omega_d$ is close to the frequency of the resonators $\omega_r$. The bandwidth of circulator behavior is set by $\kappa$ in the configuration considered here. The condition $\kappa = \kappa'/2$ is required to achieve 100\% transmission, and zero reflection at the input port.

\subsection{Tunable band structure}
Incorporating coupler circuits into larger arrays of resonators is useful for several reasons. As mentioned before, it may provide access to strongly correlated states of interacting photons with broken time-reversal symmetry. However, the usefulness of coupler circuits, is not limited to the interacting case. When leaving time-reversal symmetry intact, coupler circuits enables one to vary the (real-valued) photon hopping strength in situ and thus to systematically explore the quantum phase transition between a photonic superfluid and Mott insulator  \cite{greentree_quantum_2006,hartmann_strongly_2006,angelakis_photon-blockade-induced_2007,rossini_mott-insulating_2007,rossini_photon_2008,na_strongly_2008-1,aichhorn_quantum_2008,zhao_insulator_2008,koch_superfluidmott-insulator_2009,schmidt_strong_2009-1,schmidt_excitations_2010-1}. Finally, when breaking time-reversal symmetry both magnitudes and phases of the photon hopping elements become tunable, which can make the photonic band structure \emph{tunable} as we will show now.

\begin{figure}
	\centering
		\includegraphics[width=1.0\columnwidth]{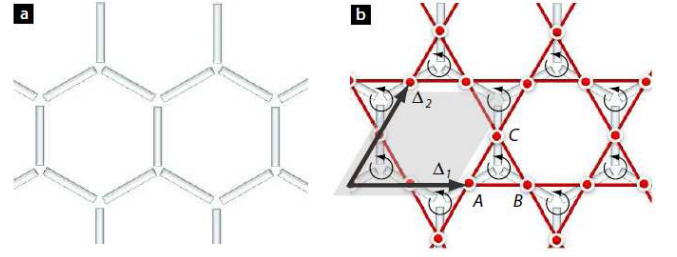}
	\caption{(a) Using three-resonator junctions one obtains a photon lattice with uniform hopping, and  the resonators (depicted as rectangles) form a regular honeycomb pattern. (b) The corresponding photon lattice is the Kagome lattice, a hexagonal Bravais lattice (primitive vectors $\pmb{\Delta}_1$, $\pmb{\Delta}_2$) with three atoms $A$, $B$, and $C$ in the primitive unit cell (parallelogram shaded in gray). Adding coupler circuits in the junctions breaks time-reversal symmetry and introduces a phase factor $e^{\pm i\varphi}$ in the photon hopping elements, where the sign depends on whether photons are transferred with or against the sense of rotation (circular arrows). \label{fig:kagome}}
\end{figure}

We consider a two-dimensional resonator array with uniform photon hopping strength. With the circuit QED realization in mind (see Section \ref{sec:realization} for details), resonators may be imagined as coplanar waveguides and uniform coupling is readily achieved by using junctions composed of three resonators at 120$^\circ$ angles. In this case, the coplanar waveguide resonators form a regular honeycomb pattern as shown in Fig.\ \ref{fig:kagome}(a). Each resonator, depicted as a rectangle, represents a single lattice site. Thus, marking the center of each resonator as a lattice site and connecting nearest neighbor sites, one finds that the photon lattice is a Kagome lattice \cite{mekata_kagome:story_2003}, see Fig.\ \ref{fig:kagome}(b).

We briefly note that, due to its novel properties and physical realizations, the Kagome lattice has played an important role in various contexts of strongly correlated systems and frustrated spin systems. Ferromagnetic and anti-ferromagnetic Ising \cite{syzi_statistics_1951,kan_antiferromagnetism.kagom_1953,wolf_ising_1988} and Heisenberg \cite{lecheminant_order_1997,waldtmann_first_1998,balents_fractionalization_2002} models have been studied on the Kagome lattice. For the Hubbard model, the Kagome lattice is known to lead to flat-band magnetism \cite{mielke__1991,mielke__1992,mielke_ferromagnetism_1993}. The possibility to create optical Kagome lattices has also created interest in exploring this physics with ultracold atoms \cite{santos_atomic_2004,isakov_hard-core_2006}. Even more recently, the (fermionic) Hubbard model on the Kagome lattice has been revisited and shown to give rise to interaction-induced topological phases \cite{guo_topological_2009,wen_interaction-driven_2010}.
Here, we show that even in the absence of interactions, the Kagome lattice displays an interesting tunable band structure when time-reversal symmetry is broken. The lattice particles (for us, photons) can assume eigenstates localized on only a few sites, giving rise to flat bands. Tuning the phase of the photon hopping makes it possible to modify the Kagome band structure and to switch the flat band to the top, middle or bottom band at will.

To demonstrate this, we consider the tight-binding model of the Kagome lattice with nearest-neighbor coupling. The Kagome lattice is generated by a hexagonal Bravais lattice with primitive vectors $\pmb{\Delta}_1=a(1,0)$ and $\pmb{\Delta}_2=\frac{a}{2}(1,\sqrt{3})$. The primitive cell contains three sites located at $\vc{r}_0=0$ ($A$), $\vc{r}_1=\pmb{\Delta}_1/2$ ($B$), and $\vc{r}_2=\pmb{\Delta}_2/2$ ($C$), where positions are expressed relative to the origin of the primitive cell. The corresponding tight-binding Hamiltonian is
\begin{widetext}
\begin{align}\label{tbham}
H=&\omega\sum_{n,m}\bigg[A_{nm}^\dag A_{nm} + B_{nm}^\dag B_{nm} + C_{nm}^\dag C_{nm}\bigg]\\\nonumber
&+t\sum_{m,n}\bigg[C_{nm}^\dag A_{nm} + B_{nm}^\dag C_{nm} + A_{nm}^\dag B_{nm} 
+C_{n,m-1}^\dag A_{nm} + B_{n-1,m+1}^\dag C_{nm} + A_{n+1,m}^\dag B_{nm}\bigg] + \text{H.c.},
\end{align}
where we have already accounted for the fact that coupler circuits may introduce photon hopping with a complex phase factor, $t=\abs{t}e^{i \varphi}$. Working in reciprocal space, we find that the dispersion $\epsilon_s(\vc{k})$ of the three bands $s=1,2,3$ is obtained from diagonalization of the following $3\times 3$-matrix:
\be
H=\sum_\vc{k}
\left(
\begin{array}{ccc}
A_\vc{k}^\dag & B_\vc{k}^\dag  & C_\vc{k}^\dag
\end{array}
\right)
\left(
\begin{array}{ccc}
\omega & 2t^*\cos(\vc{k}\cdot\pmb{\Delta}_1/2) & 2t \cos(\vc{k}\cdot\pmb{\Delta}_2/2) \\
2t\cos(\vc{k}\cdot\pmb{\Delta}_1/2) & \omega & 2t^* \cos[\vc{k}\cdot(\pmb{\Delta}_2-\pmb{\Delta}_1)/2] \\
2t^*\cos(\vc{k}\cdot\pmb{\Delta}_2/2) & 2t \cos[\vc{k}\cdot(\pmb{\Delta}_1-\pmb{\Delta}_2)/2] & \omega 
\end{array}
\right)
\left(
\begin{array}{c}
A_\vc{k} \\ B_\vc{k} \\C_\vc{k}
\end{array}
\right).
\ee
Compact analytical expressions for the band structure can be obtained for the special values $\varphi=0$, $\varphi=\pi/6$ and $\varphi=\pi/3$:
\begin{align}\label{edisp}
&\varphi=0:\quad &\epsilon_1(\vc{k})=\omega-2t,\qquad
&\epsilon_{2,3}(\vc{k})=\omega+t\pm\abs{t}\sqrt{1+8\cos[\textstyle\frac{1}{2}\vc{k}\cdot\pmb{\Delta}_1]\cos[\textstyle\frac{1}{2}\vc{k}\cdot\pmb{\Delta}_2]\cos[\textstyle\frac{1}{2}\vc{k}\cdot(\pmb{\Delta}_1-\pmb{\Delta}_2)]},\\
&\varphi=\pi/6: &\epsilon_{2}(\vc{k})=\omega,\qquad
&\epsilon_{1,3}(\vc{k})=\omega \pm2\abs{t}\sqrt{1+2\cos[\textstyle\frac{1}{2}\vc{k}\cdot\pmb{\Delta}_1]\cos[\textstyle\frac{1}{2}\vc{k}\cdot\pmb{\Delta}_2]\cos[\textstyle\frac{1}{2}\vc{k}\cdot(\pmb{\Delta}_1-\pmb{\Delta}_2)]}.
\end{align}
\end{widetext}
The case $\varphi=\pi/3$ can be show to be equivalent to $\varphi=\pi$ and is obtained from Eq.\ \eqref{edisp} by switching the sign of $t$. The band structure for $\varphi\in\frac{\pi}{3}\ZZ$ is thus familiar from previous work, see e.g.\ Ref.\ \cite{nishino_flat_2003}. Some of the results on the tight-banding band structure with broken time-reversal symmetry and an evaluation of the bands' Chern numbers have also recently been published in \cite{green_isolated_2010}.

\begin{figure*}
	\centering
		\includegraphics[width=0.9\textwidth]{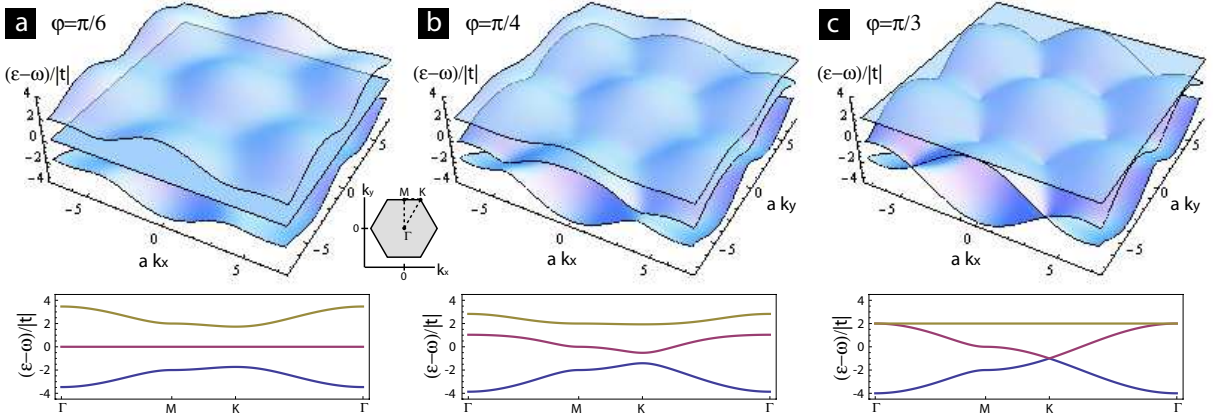}
	\caption{Band structure of the Kagome lattice with complex hopping elements $t=\abs{t}e^{i\varphi}$ for (a) $\varphi=\pi/6$, (b) $\varphi=\pi/4$, and (c) $\varphi=\pi/3$. In the top panels, the dispersion $(\epsilon_s-\omega)$ of the three bands $s=1,2,3$ is plotted in units of $\abs{t}$. The first Brillouin zone corresponds to the hexagon centered at $\vc{k}=0$. The bottom panels show cuts of the dispersion along axes of high symmetry (see inset). For phases $\varphi\in\frac{\pi}{6}\ZZ$, the band structure exhibits flat bands. The position of the flat band can be switched from (a) middle to (c) top to bottom [for $\varphi=0$, obtained from (c) by reflecting all bands at $(\epsilon-\omega)=0$] by varying the phase $\varphi$.  \label{fig:banddispersion}}
\end{figure*}

The Kagome band structure for zero and nonzero $\varphi$ is depicted in Fig.\ \ref{fig:banddispersion}. [For a beautiful discussion of the Dirac points in the band structure, visible in Fig.\ \ref{fig:banddispersion}(c), and the lifting of	the degeneracy by breaking time-reversal symmetry, see Ref.\ \cite{haldane_possible_2008}.] The characteristic flat bands occur exactly when $\varphi\in\frac{\pi}{6}\ZZ$ and depending on the specific phase, the flat band takes the role of the bottom or top band ($\varphi=0$ and $\varphi=\pi/3$), or that of the middle band ($\varphi=\pi/6$ and $\varphi=\pi/2$). We note that phase values $\varphi\notin[0,2\pi/3)$ can always be mapped back into this interval via gauge transformations. 

Band flatness and the corresponding zero group velocity are directly related to the existence of localized states \cite{mielke_ferromagnetism_1993,nishino_flat_2003}. First, consider the phase values $\varphi=\pi/6,\pi/2$ where the middle band is flat. For periodic boundary conditions with a total of $N$ primitive cells, the flat band corresponds to $N$ energy-degenerate states. This degenerate subspace is spanned by the localized hexagon states $\ket{\psi_n}$, where $\ket{\psi_n}$ is defined as the eigenstate localized on the $n$-th hexagon in the Kagome lattice with wavefunction amplitudes 
\be
\bket{jn}{\psi_n}=(-1)^j e^{ i j \pi/3}
\ee
 on the six consecutive sites $j=0,1,\ldots,5$ of the hexagon. (Note that the $\ket{\psi_n}$ states are linearly independent but non-orthogonal.)
 
When the flat band is the top (or bottom) band, the situation is slightly more complicated since the flat band touches the middle band at the $\vc{k}=0$ point and the degenerate subspace is $(N+1)$-dimensional. The localized hexagon states with amplitudes $\bket{jn}{\psi_n}=(-1)^j$ are eigenstates, but are not linearly independent since their $\vc{k}=0$ superposition $\sum_n\ket{\psi_n}$ is identically zero. The localized states can be shown to span an $(N-1)$-dimensional subspace, and the missing two $\vc{k}=0$ states  are obtained as
\begin{align}
\ket{\vc{k}=0;1}& = \frac{1}{\sqrt{2N} }\sum_{mn}(A_{mn}^\dag-B_{mn}^\dag)\ket{0},\\
\ket{\vc{k}=0;2}& = \frac{1}{\sqrt{2N}}\sum_{mn}(A_{mn}^\dag-C_{mn}^\dag)\ket{0}.
\end{align}
The existence of localized photon states and the tunability of its band structure make the Kagome lattice with variable phase factors an interesting system for future experiments. Further theoretical studies will address the interesting question of strongly correlated states induced  by photon interactions, which are expected to be non-perturbative in the presence of the flat band degeneracies of the Kagome lattice.

\section{Physical realization in the circuit-QED architecture\label{sec:realization}}
Following the general discussion of broken time-reversal symmetry in photon lattices, we now turn to a concrete proposal on how to realize this physics in the circuit-QED architecture. The essential idea is to insert superconducting circuits into the junctions between resonators. These circuits then serve as coupling elements that transfer photons from one resonator to another and may break time-reversal symmetry.

Our analysis will be organized into three subsections. In the first one, Section \ref{sec:eigenmodes}, we present the appropriate tools for modeling a transmission-line resonator capacitively coupled to arbitrary circuits at its two ends. We show how to systematically obtain the exact eigenmodes of the resonator when it is coupled to arbitrary circuits at its two ends. These exact eigenmodes are then utilized in Subsection \ref{sec:fullarray} to obtain the full Hamiltonian of a resonator array including coupling circuits. Circuit quantization \cite{devoret_quantum_1997} allows one to switch to the quantum mechanical description of the full system.

Notation in this section is heavy due to different types of objects (resonators, Josephson rings, etc.) that need to be enumerated, and we have made every effort to be consistent in our naming of indices. For reference, the different labels are summarized in Table \ref{tab}.

\begin{table}
	\centering
		\begin{tabular}{ll}
		\hline\hline
		index &  meaning\\\hline
		$i\in\{1,\ldots,N\}$ & index decomposing resonator into LC elements\\
		$j\in\ZZ$       & Josephson ring index\\
		$k\in\NN$ & excitation index for Josephson ring\\
		$\lambda\in\ZZ$ & resonator index\\		
		$\mu,\, \mu_{\lambda j}\in\{1,2,3\}$  & component of ring $j$ coupling to resonator $\lambda$\\
		$\nu\in\NN$     & resonator mode index\\
\hline\hline			
		\end{tabular}
		\caption{Summary of conventions for indices and their meanings, as used throughout Sections \ref{sec:realization} and \ref{sec:photonH}.\label{tab}}
\end{table}

\subsection{Exact resonator eigenmodes in the presence of coupling \label{sec:eigenmodes}}
\begin{figure}
	\centering
		\includegraphics[width=0.75\columnwidth]{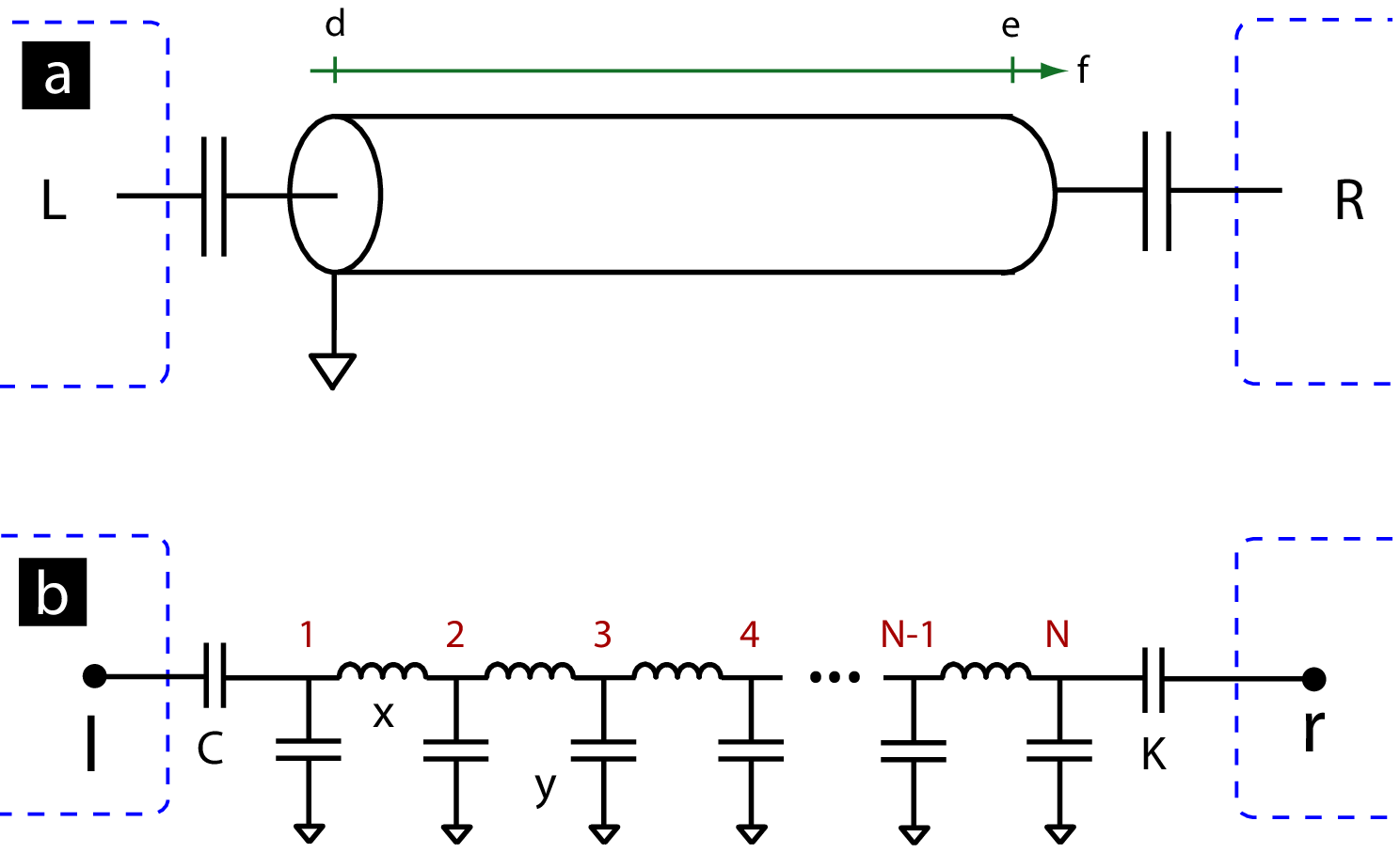}
	\caption{(a) Transmission-line resonator attached through capacitors $C_L$ and $C_R$ at the left and right ends to arbitrary circuits. Panel (b) shows the dissection of the transmission line (capacitance and inductance per unit length denoted by $c$ and $\ell$) into a series of LC circuits. The generalized flux variables adjacent to the resonator are given by $\phi_L$ and $\phi_R$.\label{fig:TL}}
\end{figure}

We consider a system consisting of a transmission line coupled capacitively at its two ends to circuits described by Lagrangians $\mathcal{L}_{L,R}'$. The general configuration is depicted in Fig.\ \ref{fig:TL}. The Lagrangian of the full system can be cast into the form
\begin{widetext}
\begin{align}\label{trline}
&\mathcal{L}=\mathcal{L}_L'+ \mathcal{L}_R'+
\frac{1}{2}C_L(\dot{\phi}_1-\dot{\phi}_L)^2 +
\frac{1}{2}C_R(\dot{\phi}_N-\dot{\phi}_R)^2+
\frac{1}{2}\sum_{i=1}^N c\, dz\, \dot{\phi}_i^2 -
\frac{1}{2\ell dz}\sum_{i=2}^N(\phi_i-\phi_{i-1})^2\\\nonumber
&= \underbrace{\sum_{\alpha=L,R}(\mathcal{L}_\alpha'+{\textstyle\frac{1}{2}}C_\alpha\dot{\phi}_\alpha^2)}_{\mathcal{L}_L+\mathcal{L}_R} 
\underbrace{-C_L\dot{\phi}_1\dot{\phi}_L
-C_R\dot{\phi}_N\dot{\phi}_R}_{\mathcal{L}_\text{int}}
+\underbrace{{\textstyle\frac{1}{2}}C_L\dot{\phi}_1^2
+{\textstyle\frac{1}{2}}C_R\dot{\phi}_N^2
+
\frac{1}{2}\sum_{i=1}^N c\, dz\, \dot{\phi}_i^2 -
\frac{1}{2\ell dz}\sum_{i=2}^N(\phi_i-\phi_{i-1})^2}_{\mathcal{L}_\text{tl}},
\end{align}
\end{widetext}
where the contributions $\mathcal{L}_{L,R}$ describe the circuits to the left and right (now including an additional capacitive contribution $\sim C_{L,R}$ due to the coupling), and $\mathcal{L}_\text{tl}$ the transmission-line resonator, modeled by an array of LC oscillators with capacitances $c\,dz$ and inductances $\ell\, dz$, where $c$, $\ell$ denote the capacitance  and inductance per unit length. The capacitive interaction between resonator and attached circuits is denoted $\mathcal{L}_\text{int}$.

It is useful to rewrite the transmission-line Lagrangian in compact matrix notation, 
\be\label{Ltl}
\mathcal{L}_\text{tl}=\frac{1}{2}\pmb{\dot{\phi}}^\top\mathsf{T}\pmb{\dot{\phi}} 
-\frac{1}{2}\pmb{\phi}^\top\mathsf{V}\pmb{\phi},
\ee
with $\pmb{\phi}^\top=(\phi_1,\ldots,\phi_N)$, 
\be
(\mathsf{T})_{ii'}=\delta_{ii'}\left( c \,dz\,+C_L\delta_{i1} + C_R \delta_{iN} \right),
\ee
and
\be
\mathsf{V}=\frac{1}{\ell\,dz}\left(
\begin{array}{rrrrrr}
1 & -1\\
-1 & 2 & -1\\
   & -1 & 2& -1 \\
   &    &  & \ddots\\
   &    &  &  -1 & 2 & -1\\
   &    &  &     & -1 & 1
\end{array}\right).
\ee
Generally, the eigenmodes $\pmb{\phi}=\zeta_\nu\vc{a}_\nu e^{-i\omega_\nu t}$ of the transmission line resonator are found by solving the generalized eigenproblem $\mathsf{V}\vc{a}_\nu = \omega_\nu^2\mathsf{T}\vc{a}_\nu$ with normalization condition $\vc{a}_\nu^\top\mathsf{T}\vc{a}_\mu=\delta_{\mu\nu}$ \cite{goldstein_classical_2001}. In the new coordinates $\pmb{\phi}=\sum_i\phi_i \vc{e}_i = \sum_\nu \zeta_\nu \vc{a}_\nu$ the resonator Lagrangian takes the simple form
\be
\mathcal{L}_\text{tl}=\frac{1}{2}\sum_\nu( \dot{\zeta}_\nu^2-\omega_\nu^2\zeta_\nu^2),
\ee
where $\nu=0,1,2,\ldots$ enumerates the resonator modes.

In our case, the kinetic matrix $\mathsf{T}$ is readily invertible. This allows us to further simplify the problem: instead of a generalized eigenproblem, we only need to solve the ordinary eigenvalue problem
\be\label{eigenval}
\mathsf{T}^{-1}\mathsf{V}\vc{a}_\nu=\omega_\nu^2\vc{a}_\nu,
\ee
 with eigenvector normalization again given by $\vc{a}_\nu^\top \mathsf{T}\vc{a}_\mu = \delta_{\mu\nu}$. Explicitly, the matrix on the left-hand side of Eq.\ \eqref{eigenval} reads
\begin{align}
&\mathsf{T}^{-1}\mathsf{V}\\\nonumber
&=\frac{1}{\ell c (dz)^2}\left(
\begin{array}{cccc}
\frac{c\, dz}{C_L+c\, dz} & -\frac{c\, dz}{C_L+c \,dz}\\
-1 & 2 & -1\\
      &  & \ddots\\
        &  -1 & 2 & -1\\
        &     & -\frac{c\, dz}{C_R+c\, dz} & \frac{c\, dz}{C_R+c\, dz}
\end{array}\right).
\end{align}
In the continuum limit, where the number of LC elements $N$ is sent to infinity and the length of the resonator $L=N\, dz$ is kept constant, the discrete mode vector $\vc{a}_\nu$ turns into the continuous mode function $\varphi_\nu(z)$. From the rows $i=2,\ldots,(N-1)$ of the matrix equation \eqref{eigenval}, one extracts the second-order differential equation
\be\label{ode}
\frac{d^2\varphi_\nu}{dz^2} = - (\omega_\nu\sqrt{\ell c})^2\varphi_\nu(z).
\ee
The rows $i=1$ and $i=N$ yield the homogeneous boundary conditions
\begin{align}
-\frac{d\varphi_\nu}{dz}\bigg|_{z=0} &= \ell C_L \omega_\nu^2\,\varphi_\nu\bigg|_{z=0},\\
\frac{d\varphi_\nu}{dz}\bigg|_{z=L} &= \ell C_R \omega_\nu^2\,\varphi_\nu\bigg|_{z=L}.
\end{align}
Finally, the orthonormalization condition turns into
\begin{align}
&C_L\varphi_\nu\varphi_\mu\bigg|_{z=0}
+C_R\varphi_\nu\varphi_\mu\bigg|_{z=L}
+c\int_0^L dz\, \varphi_\nu(z)\varphi_\mu(z)\nonumber\\
&= \delta_{\mu\nu}\label{norm}.
\end{align}
Together, Equations \eqref{ode}--\eqref{norm} form a Sturm-Liouville problem \footnote{We note that the weight function in the orthonormalization condition \eqref{norm} is slightly anomalous for a Sturm-Liouville problem, as it contains Dirac delta functions.} which determines the sinusoidal mode functions 
\be
\varphi_\nu(z)=A \cos(\omega_\nu\sqrt{\ell c}\, z) + B \cos(\omega_\nu\sqrt{\ell c}\, z)
\ee
and the corresponding mode frequencies $\omega_\nu$. The frequencies are obtained as solutions of the transcendental equation
\be\label{transc}
\tan \bar\omega = -\frac{(\chi_L+\chi_R)\bar{\omega}}{1-\chi_L\chi_R\bar{\omega}^2},
\ee
where $\bar\omega=\omega\sqrt{\ell c}L$ and $\chi_\alpha=C_\alpha/(cL)$. 
We emphasize that the treatment presented in this section has been exact and no assumptions have been made regarding the strength of the coupling between the resonator and the left and right circuits. In total, the exact Lagrangian \eqref{trline} can be written in terms of transmission-line eigenmodes as
\be
\mathcal{L} = \sum_{\alpha=L,R}\mathcal{L}_\alpha
+\frac{1}{2}\sum_\nu(\dot{\zeta}_\nu^2-\omega_\nu^2\zeta_\nu^2) - \sum_\alpha C_\alpha \dot{\phi}_\alpha\sum_\nu \dot{\zeta}_\nu\varphi_\nu(z_\alpha).
\ee

\subsection{Model for array of resonators and coupling elements\label{sec:fullarray}}
\begin{figure}
	\centering
		\includegraphics[width=0.95\columnwidth]{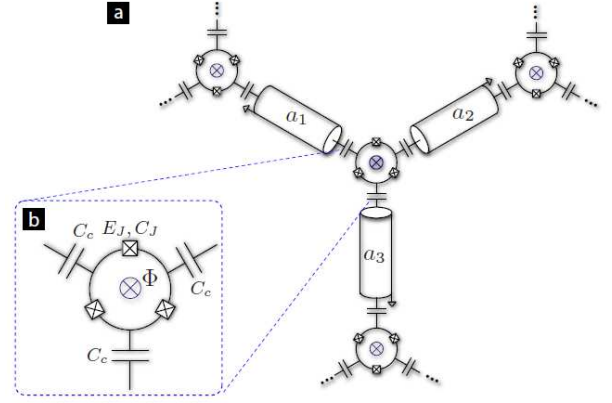}
	\caption{(Color online) (a) Array consisting of transmission-line resonators and coupling circuits in the junctions between resonators.  The coupling circuits, attached to the resonators by capacitors $C_c$, are  Josephson rings, see panel (b). They consist of a superconducting ring interrupted by three identical Josephson junctions with Josephson energy $E_J$ and junction capacitance $C_J$. By applying an external magnetic field perpendicular to the plane, the loops may additionally be threaded by a magnetic flux $\Phi$.\label{fig:array}}
\end{figure}

For the derivation of the Hamiltonian describing an array of resonators coupled by identical superconducting circuits at resonator junctions (see Fig.\ \ref{fig:array}), we consider the regime of weak coupling, as realized in the majority of circuit QED experiments. Specifically, we will assume that the coupling capacitors $C_c$ (here, $C_c=C_L=C_R$), connecting transmission-line resonators and coupling circuits, are small compared to the total capacitance of the resonator, i.e.\ $C_c\ll cL$. In this weak-coupling regime, the Hamiltonian takes a particularly simple and intuitive form as we shall demonstrate in the following. 

Quite generally, the Lagrangian of the array can be written as
\be
\mathcal{L} = \sum_\lambda \mathcal{L}_{\text{tl},\lambda} 
+ \sum_j \mathcal{L}_{\text{ri},j} + \sum_{\lambda,j} \mathcal{L}_{\text{int},\lambda,j},
\ee
where the terms describe the transmission-line resonators (``tl"), the ring circuits embedded in the resonator junctions (``ri"), and the interaction between them (``int"), respectively. As shown in the previous subsection, the resonator Lagrangian can be written in terms of eigenmodes $\nu=0,1,\ldots$ as
\be
\mathcal{L}_{\text{tl},\lambda}=\frac{1}{2}\sum_\nu(\dot{\zeta}_{\lambda\nu}^2-\omega_\nu^2 \zeta_{\lambda\nu}^2).
\ee 
We note that for small ratios $C_c/(cL)$ the transcendental equation \eqref{transc} can be solved approximately, and the lowest modes are given by $\omega_\nu\approx\nu\omega_o$. Here, the fundamental frequency corresponds to the $\lambda/2$ resonance and is given by $\omega_o/2\pi=(2\sqrt{\ell c}L)^{-1}$. %is determined by the resonator's capacitance per unit length $c$, its inductance per unit length $\ell$, and its total length $L$.

The coupling elements, which will be realized as small superconducting circuits [Fig.\ \ref{fig:array}(b)]  and discussed in more detail below, have the generic Lagrangian
\be
\mathcal{L}_{\text{ri},j} = \frac{1}{2}\pmb{\dot{\phi}}_j^\top \mathsf{C} \dot{\pmb{\phi}}_j - V(\pmb{\phi}_j,\Phi),
\ee
where $\mathsf{C}$ is the circuit's capacitance matrix and $V$ collects all inductive contributions of the circuit, including the effect of a magnetic flux $\Phi$ applied to the rings. Finally, the capacitive interaction between coupling circuits and resonators is given by
\be
\mathcal{L}_{\text{int},\lambda,j} = -\mathfrak{m}_{\lambda j}C_c (\vc{e}_{\mu_{\lambda j}}^\top\dot{\pmb{\phi}}_{j}) \sum_\nu \dot{\zeta}_{\lambda\nu} \varphi_\nu(z_{\lambda j}).
\ee
Here, $\mathfrak{m}_{\lambda j}$ plays the role of an adjacency matrix which contains all information about which resonators are coupled to which rings. It is hence defined as
\be
\mathfrak{m}_{\lambda j}=\begin{cases} 
1 & \text{if resonator } \lambda \text{ couples to ring } j,\\
0 & \text{otherwise.}
\end{cases}
\ee
Since each ring consists of three superconducting islands, we further define a component function $\mu_{\lambda j}\in\{1,2,3\}$ 
which selects the individual degree of freedom involved in the coupling between ring $j$ and resonator $\lambda$; $\vc{e}_{\mu_{\lambda_j}}$ is the corresponding three-component unit vector.  The coupling capacitors (assumed identical across the array) are denoted by $C_c$, and $z_{\lambda j}=0,L$ gives the $z$ variable entering the resonator mode function $\varphi_\nu$ [as defined in the previous subsection, Eqs.\ \eqref{ode}--\eqref{norm}].

To put the circuit and resonator variables on equal footing, it is convenient to temporarily rescale the circuit variables $\dot{\pmb{\phi}}_j \to C_o^{-1/2}\dot{\vc{F}}_\alpha$ so that $\dot\zeta_{\lambda\nu}$ and 
$\dot{\vc{F}}_j$ have identical dimensions. $C_o$ has dimensions of a capacitance, and its magnitude is chosen such that the nonzero entries in the rescaled capacitance matrix $\mathsf{K}_\alpha=C_o^{-1}\mathsf{C}_\alpha$ are of order unity. 

With these preparations it is possible to obtain an approximate expression for the Hamiltonian describing the resonator array coupled via Josephson rings. First, the conjugate momenta are obtained as
\begin{align}\nonumber
q_{\lambda\nu} &= \frac{\partial\mathcal{L}}{\partial\dot{\zeta}_{\lambda\nu}} =
\dot{\zeta}_{\lambda\nu} 
-\sum_j\mathfrak{m}_{\lambda j}\frac{C_c}{\sqrt{C_o}} (\vc{e}_{\mu_{\lambda j}}^\top \dot{\vc{F}}_j) \varphi_\nu(z_{\lambda j}),\\
\bar{\vc{Q}}_j  &= \frac{\partial\mathcal{L}}{\partial\dot{\vc{F}}_j} =
\mathsf{K}\dot{\vc{F}}_j -\sum_{\lambda,\nu}\mathfrak{m}_{\lambda j}\frac{C_c}{\sqrt{C_o}} \vc{e}_{\mu_{\lambda j}} \dot{\zeta}_{\lambda\nu} \varphi_\nu(z_{\lambda j}).\label{rel1}
\end{align}
The coupling terms on the right-hand side of the last two equations
are small in the weak-coupling limit, $C_c/\sqrt{C_o\, cL}\ll1$ valid whenever $cL\gg C_c,C_o$ \footnote{To extract the correct scaling of the coupling terms, it is important to note that the mode functions obey $\varphi_\nu(z_{\lambda j})\sim1/\sqrt{cL}$ according to their normalization.}. The inverse of Eqs.\ \eqref{rel1}, required for the Legendre transform, can then be approximated by
\begin{align}\nonumber
\dot{\zeta}_{\lambda\nu} &\approx q_{\lambda\nu}
 +\sum_j\mathfrak{m}_{\lambda j}\frac{C_c}{\sqrt{C_o}} (\vc{e}_{\mu_{\lambda j}}^\top \mathsf{K}^{-1}\bar{\vc{Q}}_j) \varphi_\nu(z_{\lambda j}),\\
\dot{\vc{F}}_j  &\approx 
\mathsf{K}^{-1}\bar{\vc{Q}}_j -\sum_{\lambda,\nu}\mathfrak{m}_{\lambda j}\frac{C_c}{\sqrt{C_o}}(\mathsf{K}^{-1} \vc{e}_{\mu_{\lambda j}}) \dot{\zeta}_{\lambda\nu} \varphi_\nu(z_{\lambda j}).
\end{align}
In these last equations, we have retained the leading order, and corrections are of the order of $\mathcal{O}(C_c^2/[C_o\, cL])$.
As a result, the weak-coupling Hamiltonian can be written in the form
\be
H=\sum_\lambda H_{\text{tl},\lambda} + \sum_j H_{\text{ri},j} + \sum_{\lambda,j} H_{\text{int},\lambda,j},
\ee
with 
\be
H_{\text{tl},\lambda}=\frac{1}{2}\sum_\nu(q_{\lambda\nu}^2+\omega_\nu^2\zeta_{\lambda\nu}^2)=\sum_\nu\omega_\nu(a^\dag_{\lambda\nu} a_{\lambda\nu}+\frac{1}{2})
\ee
and
\be
H_{\text{ri},j}= \frac{1}{2}\vc{Q}_j^\top \mathsf{C}^{-1} \vc{Q}_j + V(\pmb{\phi}_j,\Phi).
\ee
(Note that we have reverted back from our temporary rescaling and that $\vc{Q}_j=C_o^{-1/2}\bar{\vc{Q}}_j$ has proper dimensions of electric charge.)
Finally, the coupling Hamiltonian is given by
\be\label{Hcoupl}
 H_{\text{int},\lambda,j} = \mathfrak{m}_{\lambda j}C_c(\vc{e}_{\mu_{\lambda j}}^\top \mathsf{C}^{-1}\vc{Q}_j)\sum_\nu q_{\lambda\nu}\varphi_\nu(z_{\lambda j}),
\ee 
The form of the coupling Hamiltonian obtained with Eq.\ \eqref{Hcoupl} has a simple interpretation: the voltage $V_{j\mu}=\vc{e}_{\mu_{\lambda j}}^\top \mathsf{C}^{-1}\vc{Q}_j$ of coupling element $j$ (component $\mu$) is coupled by the capacitor $C_c$ to the voltage $\sum_\nu q_{\lambda\nu}\varphi_\nu(z_{\lambda j})$ at the corresponding end of resonator $\lambda$. It is important to note that in the Hamiltonian formalism, this intuitive form of the coupling is valid only in the weak-coupling limit. As soon as higher-order terms $\mathcal{O}(C_c^2/[C_o\, cL])$ are included, the coupling becomes more complicated.

\subsection{Josephson ring couplers\label{sec:Jrings}}
The coupling elements [see Fig.\ \ref{fig:array}(b)] are located in the resonator junctions and are composed of superconducting loops, each interrupted by three identical Josephson junctions. By applying an external magnetic field $\vc{B}$, each loop may be threaded by a magnetic flux $\Phi$. For reasons to be detailed below, we additionally consider the possibility of tuning the electric potential of the three superconducting islands by coupling them capacitively ($C_g$) to gate voltage sources. The Hamiltonian for one such coupling circuit is then given by
\be\label{ring-ham}
H_{\text{ri},j}=\frac{1}{2}(\vc{Q}_j-\vc{q}_j)^\top\mathsf{C}^{-1}(\vc{Q}_j-\vc{q}_j)+V(\pmb{\phi}_j,\Phi),
\ee
where the charge vector $\vc{Q}_j^\top=(Q_{j,1},Q_{j,2},Q_{j,3})$ collects the charges on nodes $\mu=1,2,$ and $3$ of Josephson ring number $j$. Similarly, $\vc{q}_j=C_g\vc{v}_j$ is composed of the corresponding offset charges. The first term thus represents the ring's charging energy and involves the inverse of the capacitance matrix
\be
\mathsf{C}=\left(
\begin{array}{rrr}
C_\Sigma & -C_J & -C_J\\
-C_J & C_\Sigma & -C_J \\
-C_J & -C_J &C_\Sigma  \\	
\end{array}\right),
\ee
built from the junction capacitances $C_J$ and the sum capacitances $C_\Sigma=2C_J+C_c+C_g$. The inductive energy contributions are given by
\be
V(\pmb{\phi}_j,\Phi)=-E_J\sum_{\mu=1}^3 \cos\bigg[ \frac{2\pi}{\Phi_0}(\phi_{j,\mu+1}-\phi_{j,\mu}-\Phi/3) \bigg],
\ee
where the $\mu$ indices, enumerating the superconducting islands within one ring $j$, are understood modulo 3, i.e. $\mu+1=4$ and $\mu=1$ are to be identified. For the following discussion, it is convenient to drop the ring index ``$j$'' and to switch to dimensionless charge and flux variables defined by $n_\mu=Q_\mu/(2e)$, $\varphi_\mu=2\pi\phi_\mu/\Phi_0$, and $\varphi=2\pi\Phi/\Phi_0$.

It is intuitively clear that the total charge $N=n_1+n_2+n_3$ on each ring is a conserved quantity. Formally, this can be confirmed by demonstrating that the total charge operator and the ring Hamiltonian commute, i.e., using the canonical commutators
$[n_\mu,e^{\pm i\varphi_{\mu'}}]=\mp\delta_{\mu\mu'}e^{\pm i\varphi_\mu}$ one verifies that $[N,H_\text{ri}]=0$ holds. 
The eigenstates of the Josephson ring Hamiltonian can consequently be written in the form $\ket{N,k}$, where $k=0,1,\ldots$ enumerates the eigenstates in the subspace of total charge $N$.

 We assume that a residual coupling of the circuit to its environment allows it to relax into its ground state $\ket{\psi_0}=\ket{N_0,0}$. Noting that the interaction Hamiltonian $H_\text{int}$ also commutes with $N$, we will assume that, for the duration of an experiment, the circuit remains in this ground state. The virtual intermediate states involved in the transfer of photons correspondingly belong to the same total charge subspace and hence can be written as $\ket{N_0,k}$.

\begin{figure}
	\centering
		\includegraphics[width=0.7\columnwidth]{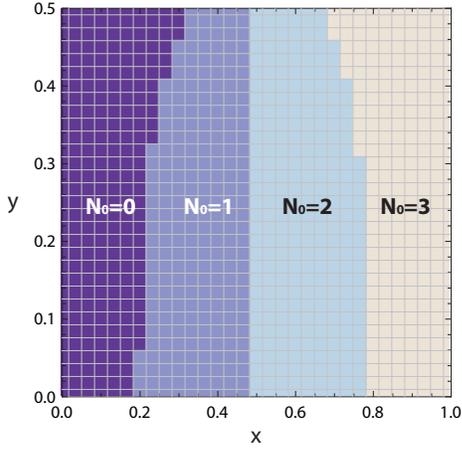}
	\caption{(Color online) Dependence of the ground state charge number $N_0$ on external magnetic flux $\Phi$ and offset charges, here for the uniform case $n_{g1}=n_{g2}=n_{g3}\equiv n_g$. As expected, $N_0$ takes on only integer values corresponding to the total number of extra Cooper pairs located on the Josephson ring. The integer-step boundaries between regions of different $N_0$ in general acquire a small finite width due to the residual coupling to the environment that allows charge relaxation. Parameters chosen for this plot: $E_J/h=10\,\text{GHz}$, $C_J=0.7\,\text{fF}$, and $C_c=5\,\text{fF}$, yielding $E_J/E_\Sigma\sim2$. \label{fig:N0figure}}
\end{figure}
 
Since, in the general case, the ring Hamiltonian is not amenable to an analytical solution, we obtain its spectrum and the charge matrix elements (required in the subsequent subsection) by numerically exact diagonalization. Our strategy is as follows: in the first step, we employ diagonalization in the charge basis to obtain the ground state $\ket{\psi_0}$ and use it to extract the total charge,
\be
N_0=\boket{\psi_0}{N}{\psi_0}.
\ee
Numerical results for this  ground state charge in a Josephson ring with realistic parameters are presented in Fig.\ \ref{fig:N0figure}. As can be inferred from the figure, $N_0$ is generally an integer-valued function of both offset charges and external magnetic flux. In the regime of strong charging effects, the dependence on flux weakens, and explicit expressions can be obtained for the boundaries between $N_0$ regions in offset-charge space [see Appendix \ref{app:N0}].

In the second step, we may then restrict ourselves to one particular subspace of total charge $N_0$. To do so, we perform a canonical transformation
\begin{align}
\varphi_1&=\varphi_1'+\varphi_3',\quad \varphi_2=\varphi_3'-\varphi_2', \quad \varphi_3=\varphi_3'\\
n_1&=n_1',\quad
n_2=-n_2',\quad
n_3=-n_1'+n_2'+n_3',
\end{align}
after which the variable $\varphi_3'$ is cyclic and the corresponding canonical momentum $n_3'=n_1+n_2+n_3=N$ is the conserved total charge. With this, the restriction of the Hamiltonian to the $N_0$ subspace can be brought into the form
\begin{widetext}
\begin{align}\label{hring2}
H_\text{ri}^{(N_0)}=&4E_{\Sigma}\bigg( n_1' - \frac{1}{2}[n_{g1}-n_{g3}+N_0]\bigg)^2
+4E_{\Sigma}\bigg( n_2' + \frac{1}{2}[n_{g2}-n_{g3}+N_0]\bigg)^2
-4E_{\Sigma} n_1' n_2'\\\nonumber
&-E_J \cos \bigg(\varphi_1'-\frac{\varphi}{3}\bigg) - E_J \cos \bigg(\varphi_2'-\frac{\varphi}{3}\bigg) - E_J \cos \bigg(\varphi_1'+\varphi_2'+\frac{\varphi}{3}\bigg).
\end{align}
\end{widetext}
Here, the charging energy $E_{\Sigma}$ has been defined such that $4E_{\Sigma}=(2e)^2(\gamma_1-\gamma_2)$, and $\gamma_{1,2}$ are reciprocal capacitances obtained in the inversion of the capacitance matrix $\mathsf{C}$, see Appendix \ref{app:cap}.
The Hamiltonian $H_\text{ri}^{(N_0)}$ has one degree of freedom less than the original ring Hamiltonian $H_\text{ri}$, and is thus more convenient for the numerical calculation of eigenenergies and charge matrix elements. 

In preparation for the next subsection where the Josephson rings will be integrated out (relying on the dispersive limit), we finally rewrite the interaction Hamiltonian in the subspace $N_0$. For the example of a single Josephson ring coupled to three resonators, the component function $\mu$ in the coupling Hamiltonian \eqref{Hcoupl} takes the simple form $\mu_{\lambda j}=j\,\delta_{\lambda,j}$. Considering only one of the low-lying modes of the resonators, we will drop the mode index ``$\nu$" from here on, and write $\omega_r$ for the (angular) resonance frequency. For the coupling Hamiltonian we then obtain
\be\label{hint2}
H_\text{int} = C_c V_\text{rms} (\vc{a}+\vc{a}^\dag)^\top  \mathsf{C}^{-1} \vc{Q},
\ee
where the vector $\vc{a}$ collects the annihilators for the three resonators $\lambda=1,2,3$, which are obtained by rewriting $q_\lambda=\sqrt{\omega_r/2}(a_\lambda+a_\lambda^\dag)$. $V_\text{rms}=\sqrt{\omega_r/2}\varphi(0)\approx\sqrt{\omega_r/cL}$ is the root-mean-square voltage in the resonators at the relevant resonator end \footnote{We note that interesting physics in such resonator arrays may also arise from the fact that the mode function generally carries a sign and, under appropriate conditions, may introduce frustration.}. Once the Hamiltonian \eqref{hint2} is restricted to the subspace of total charge $N_0$, one can show that it assumes the form
\begin{align}
H_\text{int}^{(N_0)}=&2e\beta V_\text{rms} n_1'(a_{1}-a_{3}+\text{H.c.})\nonumber\\ 
+&2e\beta V_\text{rms} n_2'(a_{3}-a_{2}+\text{H.c.})\label{int-eff}
\end{align}
with capacitance ratio $\beta=C_c(\gamma_1-\gamma_2)$. Note that here we have discarded terms of the form $\alpha(a_\lambda+a_\lambda^\dag)$ with $\alpha$ representing a c-number. Such terms merely displace the resonator mode, and can ultimately be absorbed into a redefinition of the offset charges.

\section{Effective photon lattice Hamiltonian\label{sec:photonH}}
We now turn to the crucial step of integrating out the Josephson ring elements and specifying the conditions under which the resulting photon lattice Hamiltonian breaks time reversal symmetry. The adiabatic elimination of the degrees of freedom of the coupling circuits is based on being in the dispersive regime of large energy mismatch between photonic excitations of the resonators, and excitations of the coupling circuits. Specifically, the dispersive regime is defined by the inequality $g\ll\Delta$, where $\Delta$ represents the detuning between photonic and circuit excitations and $g$ is the effective strength of their mutual coupling. For a general and systematic exposition of the adiabatic elimination technique we refer the reader to Ref.\ \cite{cohen-tannoudji_atom-photon_1998}.

Working within the rotating-wave approximation (RWA), the total number of (dressed) photons is conserved. For a given total photon number, we define $P_0$ as the projector ($P_0^2=\openone$) onto the subspace with that photon number and with all Josephson rings occupying their ground states. The effective photon lattice Hamiltonian $H_\text{ph}$
can be obtained by a canonical transformation,
\begin{align}
H_\text{ph} &= P_0 e^{iS} H e^{-iS} P_0\\\nonumber
& = \sum_\lambda H_{\text{tl},\lambda}+\frac{1}{2}P_0[iS,H_\text{int}]P_0 + \mathcal{O}(H_\text{int}^3),
\end{align}
where the generator $S$ of the transformation is chosen such that the linear coupling between rings and resonators is eliminated. To leading order in the interaction, it is given by
\be
iS = \sum_{\alpha,\alpha'} \frac{\boket{\alpha'}{H_\text{int}}{\alpha}}{E_\alpha-E_{\alpha'}}P_0 \ket{\alpha'}\bra{\alpha}P_1 -\text{H.c.}
\ee 
where $\alpha,\alpha'$ are indices for the eigenstates of $H_\text{tl}+H_\text{ri}$ in the $P_0$ subspace, and $P_1=\openone-P_0$ projects onto the complementary subspace. The main task hence consists of evaluating the contribution $\frac{1}{2}P_0[iS,H_\text{int}]P_0$ to the effective Hamiltonian. Following the arguments about charge relaxation in the previous subsection, we carry out this evaluation in the subspace with charge $N_0$, which contains the ground state of the coupling elements. 

To illustrate our procedure, we consider the simple case of three resonators attached to a single coupling element. [The generalization to a full array can be achieved by starting from Eq.\ \eqref{Hcoupl} and projecting it onto the $N_0$ charge subspace of all rings.] In RWA, the interaction Hamiltonian  \eqref{int-eff} reads
\begin{align}
H_\text{int} \,\stackrel{\text{RWA}}{=}\; &2e\beta V_\text{rms} \sum_{k>0} \bigg[
 n_{1,k}\ket{N_0,k}\bra{N_0,0}(a_{1}-a_{3})\nonumber\\
&+n_{2,k}\ket{N_0,k}\bra{N_0,0}(a_{3}-a_{2})\bigg] + \text{H.c.},
\end{align}
where $n_{\mu,k}=\boket{N_0,k}{n_\mu'}{N_0,0}$ denotes the relevant charge matrix element. It is crucial to note that the origin of photon hopping with complex-valued hopping elements is directly based on the fact that these charge matrix elements may be non-real, as we will see momentarily. A tedious but elementary calculation shows that the effective photon Hamiltonian is given by
\be
H_\text{ph} = \sum_{\lambda=1}^3 (\omega_r + \epsilon_\lambda)a_{\lambda}^\dag a_{\lambda}
+ \sum_{\lambda=1}^3 \bigg[ t_\lambda a_{\lambda}^\dag a_{\lambda+1} + \text{H.c.}\bigg],
\ee
where the index $\lambda$ in the second term is to be understood as $\lambda\bmod 3$, and where the energy shifts and photon hopping matrix elements are found to be
\begin{align}\label{eps1}
\epsilon_1&=2(\beta e V_\text{rms})^2 \sum_{k>0}\frac{|n_{1,k}|^2}{\omega_r-E_k},\\
\epsilon_2&=2(\beta e V_\text{rms})^2 \sum_{k>0}\frac{|n_{2,k}|^2}{\omega_r-E_k},\\
\epsilon_3&=2(\beta e V_\text{rms})^2 \sum_{k>0}\frac{|n_{1,k}-n_{2,k}|^2}{\omega_r-E_k},\label{eps3}\\
t_1&=2(\beta eV_\text{rms})^2 \sum_{k>0}\frac{-(n_{1,k})^*n_{2,k}}{\omega_r-E_k},\label{t1}\\
t_2&=2(\beta eV_\text{rms})^2 \sum_{k>0}\frac{(n_{1,k})^*n_{2,k}-|n_{2,k}|^2}{\omega_r-E_k},\\
t_3&=2(\beta eV_\text{rms})^2 \sum_{k>0}\frac{(n_{1,k})^*n_{2,k}-|n_{1,k}|^2}{\omega_r-E_k}.\label{t3}
\end{align}
$E_k$ denotes the energy of the $k$-th circuit excitation (measured relative to the ground state energy $E_0$). Eqs.\ \eqref{t1}--\eqref{t3} for the hopping matrix elements confirm our previous statement that the emergence of complex phase factors in the hopping is directly linked to the possibility of non-real charge matrix elements. Before investigating the conditions under which these charge matrix elements are non-real and result in breaking of time-reversal symmetry, it is useful to note that, in general the above equations will also lead to breaking of the three-fold rotation symmetry due to the energy shifts $\epsilon_\lambda$. The origin of this is, of course, the possible presence of \emph{different} offset charges on each of the three superconducting islands. 

For the present discussion, we restrict our discussion to the case where no such breaking of the three-fold symmetry occurs, and we will hence choose identical offset charges $n_{g1}=n_{g2}=n_{g3}\equiv n_g$. In the ideal case, individual superconducting islands would not need to be connected to separate gate voltage sources; instead, a global electric field perpendicular to the chip plane (e.g., by a back gate) could be applied to achieve a uniform and tunable offset charge. (This, of course, neglects the presence of random offset charges and $1/f$ charge noise which we address in Section \ref{sec:offset}.) With the threefold symmetry intact,  one concludes that  
\be\label{e123}
\epsilon_1=\epsilon_2=\epsilon_3
\ee
 must be satisfied. In other words, application of a global electric field does not lead to energy detuning between resonators.  
 
We need to be cautious though not to throw out the baby with the bath water. Clearly, fixing all offset charges to be identical is a strong restriction of parameter space and it is by no means obvious that this leaves any freedom for complex-valued matrix elements and hence time-reversal symmetry breaking on the level of the effective photon Hamiltonian. Let us thus verify that  Eq.\ \eqref{e123} when combined with Eqs.\ \eqref{eps1}--\eqref{t3} is in general compatible with complex-valued hopping elements $t_\lambda$. Given that $\epsilon_\lambda$ must take the form of Eqs.\ \eqref{eps1}--\eqref{eps3}, a sufficient condition for satisfying $\epsilon_1=\epsilon_2=\epsilon_3$ is obtained by requiring that,  for each excitation level $k$,  the charge matrix elements $n_{\mu,k}$ have equal modulus, $\abs{n_{1,k}}=\abs{n_{2,k}}$, and obey $\abs{n_{1,k}}^2=\abs{n_{1,k}-n_{2,k}}^2$. Evaluating these conditions, we find that the charge matrix elements obey
\be
n_{\mu,k}=r_k e^{i f_{\mu,k}}
\ee
with modulus $r_k\ge0$ independent of the charge index $\mu=1,2$, and phases 
\be
f_{1,k}-f_{2,k} = (\pm)_k\frac{\pi}{3}+2\pi z_k.
\ee
The latter equation must hold for all levels $k=1,2,\ldots$, but both the sign and the integer $z_k\in \ZZ$ may differ among levels. The freedom in the phase sign turns out to be \emph{crucial} for breaking time-reversal symmetry. Without the sign freedom or when truncating the system to a two-level system, the (gauge-invariant) phase sum over the three-resonator loop would always be an integer multiple of $\pi$.
Hence, as discussed in Section \ref{sec:general}, time-reversal symmetry would be intact on the level of the effective photon Hamiltonian. However, due to sign flips for higher levels $k$ and together with the different prefactors in the terms of the sum [Eqs.\ \eqref{t1}--\eqref{t3}], arbitrary gauge-invariant phases
\be\label{giphase}
\osum_{\lambda=1}^3 \varphi_{\lambda,\lambda+1} = \arg \prod_{\lambda=1}^3 t_\lambda
\ee
 can in principle be generated and time-reversal symmetry thus be broken. 
 
\begin{figure*}
	\centering
		\includegraphics[width=1.0\textwidth]{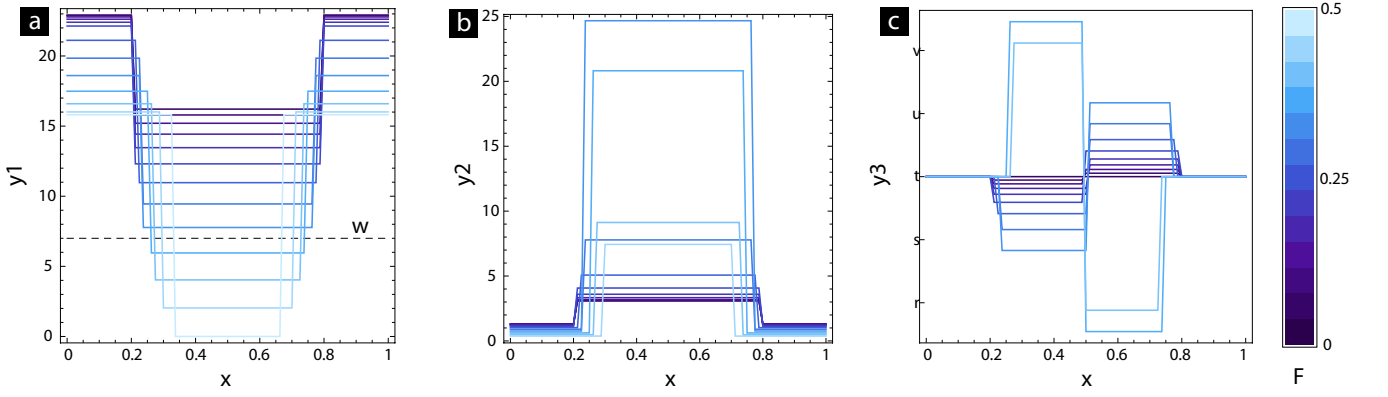}
	\caption{(Color online) Numerical results for a junction of three resonators attached to a central Josephson ring coupler. The device is tunable by varying the magnetic flux $\Phi$, see color/gray scale, and by changing the global offset charge $n_g$ as set by a constant electric field, see $x$ axes. 
	Panel (a) shows the lowest  transition frequency $\omega_{01}/2\pi$ of the Josephson ring device in comparison with the resonator frequency $\omega_r/2\pi=7\,\text{GHz}$. As one can check, the dispersive limit is maintained for the selected values of magnetic flux $\Phi$. Panel (b) displays the resulting magnitude of photon hopping strengths $\abs{t}$. The non-monotonic behavior is explained  by the crossing of the $\omega_{01}$ transition and the resonator frequency around $\Phi/\Phi_0\sim0.3$.
	Panel (c) presents the corresponding results for the gauge-invariant phase sum $\osum \varphi$ and proves the breaking of time-reversal symmetry. As expected from general considerations, time-reversal invariance remains intact at zero offset charge, and at zero magnetic flux. (Parameters as in Fig.\ \ref{fig:N0figure}, in addition: $C_c=5\,\text{fF}$, $\omega_r/2\pi=7\,\text{GHz}$ and $\sqrt{\ell/c}=50\,\Omega$.) \label{fig:numplots}}
\end{figure*}

\subsection{Numerical results for intermediate $E_J/E_\Sigma$}
Equations \eqref{t1}--\eqref{t3} allow for a direct evaluation of the essential parameters of the effective photon Hamiltonian. The most important quantity for determining whether time-reversal symmetry breaking succeeds is the gauge-invariant phase sum $\osum_\mathcal{C}\varphi$, Eq.\ \eqref{giphase}. Whenever this sum corresponds to an integer multiple of $\pi$, time-reversal symmetry is intact; for all other values it is broken. In these terms, our prime concern is to demonstrate that 
\be
\osum_\mathcal{C}\varphi\notin \pi\ZZ
\ee
can be achieved for realistic device parameters and reasonable magnitude of the photon hopping element (clearly, for hopping matrix elements with $\abs{t}=0$ the complex phase becomes arbitrary and completely meaningless).

Results from numerical diagonalization for a selected set of parameters, chosen with current fabrication capabilities and general parameter requirements in mind, are presented in Fig.\ \ref{fig:numplots}.
We find that breaking time-reversal symmetry is feasible under realistic conditions, and that the external dc electric and magnetic fields can be utilized to switch  time-reversal invariance on and off (with the electric field) and to smoothly tune the value of the gauge-invariant phase sum (with the magnetic field). 

Several comments are in order to provide an intuitive understanding of the numerical results shown in Fig.\ \ref{fig:numplots}.
We note that the excitation energies of the Josephson ring and the resulting photon hopping amplitudes and phases  exhibit a step-like dependence on the global offset charge. This is easily understood from the Josephson ring Hamiltonian, Eq.\ \eqref{hring2}: The values of the offset charges fix the total charge $N_0$. Further, in the case of identical offset charges  $n_{g1}=n_{g2}=n_{g3}$, this is the \emph{only way} the offset charges enter the Hamiltonian. By consequence, the fact that $N_0$ is an integer-valued function of $n_{g\mu}$  explains the step-wise dependence on offset charges. Only at special points where an increase in the common offset charge causes a level crossing of the two lowest states in subspaces with different total charge, the parameter $N_0$ changes discontinuously from one integer to another and thus leads to the observed steps. 

The fact that time-reversal symmetry is broken for $N_0=1,2$ (and, by means of charge periodicity, for all $N_0\bmod3=\pm 1$) and that the gauge-invariant phase sums are of opposite sign for these two cases can easily be motivated by considering the case of large charging energy. For $N_0=1$ there are three nearly degenerate states with one additional Cooper pair (the ``particle") located on one of the three islands. When $E_J$ is finite, the extra Cooper pair can start to move, becomes susceptible to the vector potential and produces
an effective phase in the photon hopping. Conversely, for $N_0=-1$ (equivalent to $N_0=2$) there are three nearly degenerate states with a Cooper pair missing (i.e., a ``hole") on one of the three islands. This results in the opposite signs of the gauge-invariant phase sums since hopping of particles involves the phase $\varphi$, whereas hopping of holes is associated with phase $-\varphi$. The case $N_0 \bmod 3 = 0$ corresponds to the particle-hole symmetric case, where the photons acquire zero synthetic gauge charge and time-reversal symmetry holds.

As we will prove below, the regime of very large $E_J/E_\Sigma$ ratios (where Josephson tunneling completely overwhelms charging effects) is inadequate for breaking time-reversal symmetry. As a result, charge noise must be expected to impose limitations on the proposed device, which we briefly address in Section \ref{sec:offset}. Future work must establish the optimal working point where $\osum_\mathcal{C}\varphi$ comfortably reaches the crucial value of $3\times\pi/6=\pi/2$ while keeping sensitivity to offset-charge fluctuations at a minimum.

\subsection{Conditions for time-reversal symmetry breaking}
First, let us establish that in the regime where Josephson tunneling dominates over charging effects, i.e., $E_J/E_\Sigma\gg1$, 
the Josephson ring \emph{fails} to break time-reversal symmetry. 
To see this, consider the ring Hamiltonian \eqref{hring2} in phase basis where $n_\mu'=i d/d\varphi_\mu'$ (we will drop primes in the following). For $E_J\gg E_\Sigma$, the Hamiltonian describes the situation of a fictitious particle with large mass in a two-dimensional potential. (Strictly speaking, the space described by the coordinates $\varphi_{1,2}$ is a torus, since the periodic boundary conditions require that $\varphi_\mu$ and $\varphi_\mu+2\pi$ be identified as the same coordinate.)  Due to the large mass, the low-energy part of the spectrum can be described by a local approximation of the two-dimensional potential at its minimum \footnote{This approximation is a multidimensional generalization of the approximation used to describe the Cooper pair box in the transmon regime \cite{koch_charge-insensitive_2007}.}, 
\be\label{vapprox}
V(\pmb{\varphi})\simeq \frac{1}{2}(\pmb{\varphi}-\pmb{\varphi}_\text{min})^\top \mathsf{M} (\pmb{\varphi}-\pmb{\varphi}_\text{min}).
\ee
Here, $\mathsf{M}$ is positive definite, and we have used the vector notation $\pmb{\varphi}=(\varphi_1,\varphi_2)$. (Note that both the curvature matrix $\mathsf{M}$ and the position of the minimum $\pmb{\varphi}_\text{min}$ still depend on the magnetic flux, which we suppress in our notation.) Once the approximation \eqref{vapprox} is employed, the periodic boundary conditions are changed into the regular boundary condition $\int_{\RR^2}d\varphi_1 d\varphi_2 \abs{\psi(\varphi_1,\varphi_2)}^2=1$. This opens the way for a gauge transformation
\be
\psi(\varphi_1,\varphi_2) = \exp(i\alpha_1\varphi_1+i\alpha_2\varphi_2)\bar{\psi}(\varphi_1,\varphi_2),
\ee
which leaves the new boundary condition unchanged. Choosing 
\be\label{alpham}
\alpha_m=(-1)^m(N_0+3n_{gm}-\textstyle\sum_{\mu=1}^3 n_{g\mu})/3,
\ee
this transformation can be used to eliminate all offset-charge related first derivatives from the Schr\"odinger equation for $\bar{\psi}$. In other words, in this gauge the fictitious particle does not ``see" a vector potential and its $\bar{\psi}$ wavefunction can be chosen entirely real-valued. This in turn reveals that all charge matrix elements can be chosen purely imaginary, and consequently all hopping elements for photons purely real-valued, $t_\mu\in\RR$ \footnote{It should be noted that this argument is not limited to the situation of identical offset charges, but is valid for arbitrary $n_{g\mu}$.}. While time-reversal symmetry is thus not broken in this regime, we emphasize that Josephson rings in the large $E_J/E_\Sigma$ regime are still very useful: they make the photon hopping strength $t_\mu$ \emph{tunable} with an external magnetic field and remain insensitive to the effects of random offset charges and $1/f$ charge noise just like the transmon qubit \cite{koch_charge-insensitive_2007,schreier_suppressing_2008}.

Closely related to the no-go statement for time-reversal symmetry breaking  with large $E_J/E_\Sigma$ ratios, one can specify two general conditions required for breaking of time-reversal symmetry. First, we note that breaking particle-hole symmetry, or equivalently, the presence of nonzero offset charges, is required. The argument for this directly follows from our previous discussion: without offset charges, all eigenfunctions of the Josephson ring Hamiltonian in phase basis can be chosen real-valued outright [i.e., without the substep of approximating the potential in Eq.\ \eqref{vapprox}]. The repetition of our arguments following Eq.\ \eqref{alpham} then again leads to the conclusion of no time-reversal symmetry breaking. For the case of identical offset charges, we can narrow down the necessary condition further: since  the Hamiltonian \eqref{hring2} remains invariant (up to an irrelevant overall constant) under the transformation $N_0\to N_0\pm3$, we find that $N_0 \bmod 3=\pm1$ is required to break time-reversal symmetry.

Second, we note that the presence of Josephson junctions is crucial in our scheme. Without them, the inductive energy would generically take the form of Eq.\ \eqref{vapprox}, and all subsequent arguments leading to the conclusion of no time-reversal symmetry breaking hold.

\subsection{Consequences of random offset charges and $1/f$ charge noise\label{sec:offset}}
It is known from experiments with superconducting charge qubits \cite{bouchiat_quantum_1998,nakamura_coherent_1999,vion_manipulatingquantum_2002,metcalfe_measuringdecoherence_2007} that the coupling of a superconducting circuit to its environment generally results in random offset charges on superconducting islands, and that these offset charges typically fluctuate as a function of time with a characteristic $1/f$ noise spectrum \cite{zorin_background_1996,kafanov,metcalfe_measuringdecoherence_2007}. This behavior will likely affect the performance of the Josephson coupler circuits proposed here, and we comment on consequences and potential solutions to this issue.

For superconducting charge qubits, the negative effects of charge noise can be cured by working with transmon qubits in the regime where Josephson tunneling dominates over charging effects  \cite{koch_charge-insensitive_2007,schreier_suppressing_2008}. This venue, however, is not available for the Josephson ring circuit when aiming at time-reversal symmetry breaking, as follows from our discussion in the previous section. While devices with large $E_J/E_\Sigma$ will be insensitive to charge noise and very useful for making photon hopping strengths tunable, the gauge-invariant phase sum around the loop will be exponentially suppressed.

\begin{figure}
	\centering
		\includegraphics[width=0.95\columnwidth]{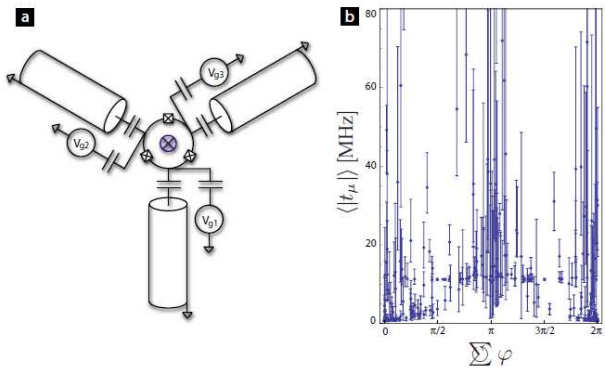}
	\caption{(Color online) (a) Josephson ring with attached voltage bias lines for cancelling random offset charges. (b) Effective photon hopping strengths and gauge-invariant phase sums for random offset charges, with $n_{g\mu}\in[0,1]$ with uniform probability distribution. Data points are placed such that their $x$ positions correspond to the gauge-invariant phase sums $\osum\varphi$ (modulo $2\pi$), and their $y$ positions display the arithmetic mean of the three photon hopping strengths $\abs{t_\mu}$. For each data point, an ``error" bar shows the spread from the minimum $\abs{t_\mu}$ to the maximum. (Device parameters used are the same as in Fig.\ \ref{fig:numplots}.)	
	\label{fig:offset}}
\end{figure}

For devices with one or maximally a few Josephson coupler circuits, it is conceivable to work with intermediate $E_J/E_\Sigma$ ratios and to couple the individual superconducting islands capacitively to voltage bias lines, see Fig.\ \ref{fig:offset}(a). This way, random offset charges can be cancelled and the device stabilized. For larger arrays, attaching individual bias lines becomes cumbersome. Random offset charges then lead to disorder in the photon hopping elements as well as in the gauge-invariant phase sums, see Fig.\ \ref{fig:offset}(b). While presence of such disorder poses interesting questions itself (compare, the recent interest in potential disorder in ultracold atom systems, see e.g. \cite{white_strongly_2009}, and localization in random magnetic fields, e.g.\ \cite{lee94,aronov94}), future studies will also aim at identifying alternative superconducting circuits for charge-noise insensitive and time-reversal symmetry breaking coupling elements.

\section{Conclusions and Outlook\label{sec:conclusions}}
In summary, we have shown that superconducting circuits based on Josephson junctions can be used to break time-reversal symmetry in arrays of on-chip microwave resonators. In the first part of our paper, we have explored how to use passive coupling elements to generate gauge-invariant phases in the lattice hopping elements, and how these phases are related to time-reversal symmetry breaking. Much of this discussion is general and can readily be transferred to lattices other than photon lattices. Our subsequent discussion has highlighted consequences and applications of breaking time-reversal symmetry in non-interacting lattices of photons, including the realization of an on-chip circulator and the achievement of a highly tunable band structure for the concrete case of a photonic Kagome lattice. We note that the existence of localized photon states on hexagons in the Kagome lattice may be of interest for photon storage in the future. These localized photon states do not necessitate the presence of a large lattice, but can rather be accessed in a single Kagome star consisting of only twelve resonators -- a setting that is well within reach of current experimental capabilities. 

The second part of our paper has addressed a concrete proposal for the realization of such passive coupling elements in the circuit QED architecture. Our presentation aimed to be pedagogical and to collect the necessary circuit quantization tools to handle an array of transmission line resonators coupled to small superconducting circuits playing the role of coupling elements. We have stated the general conditions for breaking time-reversal symmetry with a passive coupling element, including the necessity of non-linear elements (Josephson junctions), the presence of a magnetic field, and breaking of particle-hole symmetry. We have shown that an extremely simple circuit, a superconducting ring interrupted by three Josephson junctions, can be used to satisfy all the necessary requirements. For realistic device parameters, we have calculated the resulting photon hopping strengths and gauge-invariant phases as a function of external magnetic flux and global offset charge. Finally, we have identified random offset charges and charge noise as likely challenges when targeting a lattice without disorder in hopping strengths and phases. Future works will explore alternative circuits for tackling this issue, and will address the interesting question of strongly-correlated photon states with broken time-reversal symmetry, which are expected for large effective photon-photon interaction such as in the Jaynes-Cummings lattice.

\begin{acknowledgments}
This work was supported in part by Yale University via a Quantum Information and Mesoscopic Physics Fellowship (JK), by the NSF through the Yale Center for Quantum Information Physics (DMR-0653377), grants DMR-0603369 (SMG) and  DMR-0803200 (KLH), and by the Packard Foundation (AAH).
We thank Jay Gambetta, Archana Kamal, and Michel Devoret for valuable discussions.
\end{acknowledgments}

\appendix
\section{Time-reversal symmetry\label{app:treversal}}
Generally, the dynamics of a system is said to be time-reversal symmetric if for a given solution to the equations of motion, the corresponding motion-reversed evolution is a valid solution as well. In the following, we briefly compile the most important facts about time-reversal in quantum mechanics.

In quantum mechanics, symmetries manifest as maps $S$ of Hilbert space, which leave all observable probabilities invariant, i.e.\ $\abs{\bket{S\phi}{S\psi}}^2=\abs{\bket{\phi}{\psi}}^2$ for all states $\ket{\phi},\ket{\psi}$ \cite{merzbacher_quantum_1997,weinberg_quantum_2005}.   This is fulfilled if and only if $S$ is either a unitary operator, or  an operator which is anti-linear and anti-unitary \cite{wigner_gruppentheorie_1931,weinberg_quantum_2005}. While the former choice applies to discrete and continuous symmetries including rotations and parity, the latter option  must be selected for time reversal, in order to avoid energy spectra not bounded from below [see, e.g., Ref.\ \onlinecite{merzbacher_quantum_1997} for the proof of this statement]. The time-reversal operation $\Theta$ must thus be anti-linear and anti-unitary, i.e.\
\begin{align}
&\Theta(\alpha \ket{\phi} + \beta \ket{\psi})= \alpha^* \Theta\ket{\phi} + \beta^*\Theta\ket{\psi},\label{antilin}\\
&\bket{\Theta\phi}{\Theta\psi}=\bket{\psi}{\phi}\label{antiun}.
\end{align}
Once time reversal $\Theta$ has been properly defined for a specific system with Hamiltonian $H$, symmetry of that system under time reversal is signalled by the fact that $\Theta H \Theta^{-1}=H$ holds.  (For simplicity, we are excluding the case of degenerate eigenstates of $H$, for which $\Theta$ may additionally induce a rotation within the degenerate subspace.)

To define $\Theta$ explicitly, we assume that the system provides us with an observable (with non-degenerate spectrum), say $\vc{x}$, which is expected to be time-reversal invariant for physical reasons. For example, this operator may be the position operator for the location of a particle in real space; for a circuit network, it may be the operator for charge on a certain network node, which also must remain invariant under time reversal. Under these assumptions, time-reversal is expected to leave the eigenstates of $\vc{x}$ invariant, possibly up to a phase,
\be\label{Tinvariance}
\Theta \ket{\vc{x}} = e^{i\vartheta(\vc{x})}\ket{\vc{x}},
\ee
from which $\Theta \vc{x} \Theta^{-1}=\vc{x}$ immediately follows. Time-reversal symmetry thus holds if and only if there exists a phase $\vartheta(\vc{x})$ such that $\Theta H \Theta^{-1}=H$ is satisfied. We will see momentarily that the phase $\vartheta$ is intimately related to phases arising from gauge transformations.

Eq.\ \eqref{Tinvariance} has several important consequences, which we briefly gather in the following. (i) Once $\vartheta(\vc{x})$ is fixed, the action of $\Theta$ on the entire Hilbert space is uniquely defined by Eq.\ \eqref{Tinvariance}. To see this, decompose any state $\ket{\psi}$ in the position basis and invoke anti-linearity to obtain
\begin{align}
\Theta\ket{\psi}&=\int d^dx\, \Theta\bigg[\psi(\vc{x})\ket{\vc{x}}\bigg]\nonumber\\
&= \int d^dx\, e^{i\vartheta(\vc{x})}\psi^*(\vc{x})\ket{\vc{x}}.\label{theta}
\end{align}
(ii) The anti-unitarity condition, Eq.\ \eqref{antiun}, is automatically satisfied by this definition of $\Theta$.
(iii) The canonical momentum $\vc{p}$ transforms under time-reversal as 
\be\label{ptransform}
\Theta \vc{p} \Theta^{-1} = -\vc{p} + \nabla \vartheta(\vc{x}),
\ee
which can be derived using Eq.\ \eqref{Tinvariance} and the canonical commutator $[\vc{x},\vc{p}]=i$.
  
To demonstrate how the phase $\vartheta$ is determined by our gauge choice,  consider the example of a particle with mass $m$ in an external potential with Hamiltonian
$ H=\vc{p}^2/2m + V(\vc{x})$.
 Choosing $\vartheta(\vc{x})=0$, one can verify that $\Theta \vc{p} \Theta^{-1}=-\vc{p}$, and hence $\Theta H \Theta^{-1}=H$. As  expected, the problem is time-reversal symmetric. The same system can, of course, be described in a different basis, related to the original position basis by a local gauge transformation,
$ \ket{\vc{x}}\mapsto e^{i\chi(\vc{x})}\ket{\vc{x}}$.
 In the transformed basis, the Hamiltonian takes the modified form
\be
H=\frac{1}{2m}\bigg[\vc{p}+\nabla\chi(\vc{x})\bigg]^2 + V(\vc{x}).
\ee
Performing a gauge transformation cannot affect time-reversal invariance, and so $\Theta H \Theta^{-1}=H$ should hold for an appropriate choice of $\vartheta$. Indeed, using Eq.\ \eqref{ptransform}, we can construct $\vartheta$ by requiring
\be
H=\Theta H \Theta^{-1} =\frac{1}{2m}\bigg[-\vc{p}+\nabla\vartheta(\vc{x})+\nabla\chi(\vc{x})\bigg]^2 + V(\vc{x}),
\ee
which yields $\nabla\vartheta(\vc{x})+2\nabla\chi(\vc{x})=0$. As a result, the phase of the time-reversal operator is fixed by the gauge, $\vartheta(\vc{x})=-2\chi(\vc{x})$ up to an irrelevant constant. If we interpret $\vc{A}=\nabla\chi$ as a vector potential (here with zero curl), we can write 
\be
\vartheta(\vc{x})=-2\int_{\vc{x}_0}^{\vc{x}}d\vc{s}\cdot\vc{A}.
\ee
 As an immediate corollary we note that the presence of a magnetic field would manifest in a vector potential $\vc{A}$ with nonzero curl. In that case, the resulting equation $\nabla\vartheta(\vc{x})+2\vc{A}=0$ has no solutions, and hence time-reversal symmetry is broken.

 In summary, one can thus show that the following equivalences hold for the case of position and momentum operator having continuous spectra: Time-reversal symmetry is intact. $\Leftrightarrow$ There exists a phase choice for $\vartheta(\vc{x})$ such that $\Theta H \Theta^{-1}=H$ holds. $\Leftrightarrow$ There exists a  local gauge transformation that makes the Hamiltonian real-valued. $\Leftrightarrow$ The vector potential satisfies $\oint_\mathcal{C} d\vc{s}\cdot\vc{A}=0$ for any closed contour $\mathcal{C}$. (Note that non-singularity of the phase functions is implied everywhere.)

Finally, let us switch to the case of a discrete position operator, such as for a lattice Hamiltonian
\be\label{hdiscrete}
H=|t|\sum_{ j\not=k}e^{i\varphi_{jk}}a^\dag_k a_j + \sum_j \omega a^\dag_j a_j \qquad (\varphi_{kj}=-\varphi_{jk}),
\ee
describing a system of particles which can hop between lattice sites, say from $j$ to $k$, and doing so pick up a phase factor $\varphi_{jk}$.  As the analog of the continuous position basis, we use the particle number states $\ket{n_1,n_2,\ldots}$, and hence define the  time-reversal operation via
\be\label{thetan}
\Theta \ket{n_1,n_2,\ldots} = e^{i\vartheta(n_1,n_2,\ldots)} \ket{n_1,n_2,\ldots}.
\ee
For our purposes it is sufficient to consider linear functions of the form $\vartheta(n_1,n_2,\ldots)=\sum_j \vartheta_j n_j$. Invariance under time reversal is then equivalent (by definition) to the existence of phases $\vartheta_j$ such that $\Theta H \Theta^{-1}=H$ holds.

From Eq.\ \eqref{thetan} with linear $\vartheta$, one obtains the transformation law for annihilation operators, which reads
\be
\Theta a^\dag_j \Theta^{-1} = e^{i\vartheta_j} a^\dag_j.
\ee
Applying the time-reversal operation to the Hamiltonian \eqref{hdiscrete}, we thus find that invariance under time-reversal implies the existence of a set of phases $\{\vartheta_j\}$ such that
\be\label{2varphi}
\vartheta_k-\vartheta_j + 2\varphi_{kj} \in  2\pi\ZZ 
\ee
holds for all lattice indices $j,k$. (Note: once such phases $\vartheta_j$ have been found, the gauge transformation with phases $\{\vartheta_j/2\}$ makes the number-basis Hamiltonian real-valued.) The last condition \eqref{2varphi} can finally be shown to be equivalent to the requirement that 
\be
\osum_{\mathcal{C}[jk]}\varphi_{jk}\in \pi\ZZ
\ee
for all closed loops $\mathcal{C}$. The correspondences between the continuous and the discrete case are summarized in Table \ref{tab2}.

\begin{table}
	\centering
		\begin{tabular}{p{4cm}p{4cm}}
		\hline\hline
		continuous & discrete \\\hline
    $\ket{\vc{x}}$ & $\ket{n_1,n_2,\ldots}$ \\
    $\vartheta(\vc{x})$ & $\{\vartheta_j\}$\\
    $\vc{A}(\vc{x})$ & $\varphi_{kj}$ \\
    $\nabla\vartheta+2\vc{A}=0$ &  $\vartheta_k - \vartheta_j + 2\varphi_{kj} \in  2\pi\ZZ$\\
    $\int d\vc{s}\cdot\vc{A}=0$ & $\osum_{\mathcal{C}[jk]}\varphi_{jk}\in\pi\ZZ$\\
		\hline\hline	
		\end{tabular}
		\caption{Correspondences for time-reversal symmetry in continuous and discrete systems. The statements in the last two rows only hold if the system is time-reversal invariant. \label{tab2}}
\end{table}

\section{Inverse of the capacitance matrix \label{app:cap}}
For completeness, we provide explicit expressions for the inverse of the capacitance matrix $\mathsf{C}$:
\be
\mathsf{C}^{-1}=\left(
\begin{array}{rrr}
C_\Sigma & -C_J & -C_J\\
-C_J & C_\Sigma & -C_J \\
-C_J & -C_J &C_\Sigma  \\	
\end{array}\right)^{-1}=
\left(
\begin{array}{rrr}
\gamma_1 & \gamma_2 & \gamma_2\\
\gamma_2 & \gamma_1 & \gamma_2 \\
\gamma_2 & \gamma_2 & \gamma_1  \\	
\end{array}\right).
\ee
The reciprocal capacitances $\gamma_{1,2}>0$ are defined as
\begin{align}\label{gamma1}
\gamma_1&=\frac{C_\Sigma-C_J}{(C_\Sigma-2C_J)(C_\Sigma+C_J)},\\
\gamma_2&=\frac{C_J}{(C_\Sigma-2C_J)(C_\Sigma+C_J)}.
\end{align}

\section{Total charge number of the Josephson ring ground state\label{app:N0}}
\begin{figure}
	\centering
		\includegraphics[width=1.0\columnwidth]{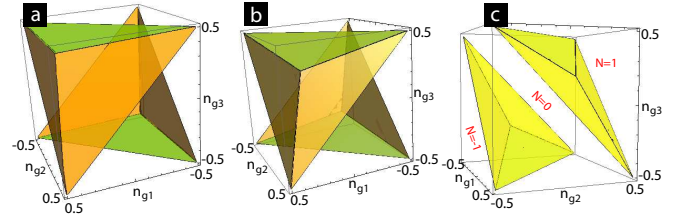}
	\caption{(Color online) Regions of fixed total charge $N=0,\pm1$ in the charge regime, as a function of the three offset charges $n_{g1}$, $n_{g2}$ and $n_{g3}$. The shape of the region boundaries depends on the charging energy ratio $E_{C}/E_{C}'$, chosen as (a) $5/4$, (b) $5/2$, and (c) $20$. Note that the coordinate axes are oriented differently in panel (c) to reveal the flatness of the boundaries for large $E_C/E_C'$. \label{fig:ntot}}
\end{figure}
In Section \ref{sec:Jrings}, we noted that the eigenstates of the Josephson ring Hamiltonian \eqref{ring-ham} can naturally be chosen as simultaneous eigenstates of the total ring charge $N=n_1+n_2+n_3$, here measured in units of the Cooper pair charge $(2e)$. For the subsequent discussion in that section, it was important to extract the total charge number of the ground state $N_0=\boket{\psi_0}{N}{\psi_0}$ for given offset charges $\vc{n}_g=(n_{g1},n_{g2},n_{g3})$ and model parameters. 
While numerical diagonalization of the Hamiltonian \eqref{ring-ham} allows the direct calculation of $N_0$, it is useful to first understand the general structure of $N_0$.

Our starting point is the ring Hamiltonian, written in terms of dimensionless charge numbers $\vc{n}=(n_1,n_2,n_3)$ and phase difference $\varphi_j$,
\begin{align}
H_\text{ri} = &4E_C (\vc{n}-\vc{n}_g)^\top \mathsf{M} (\vc{n}-\vc{n}_g)\\\nonumber
&-E_J\sum_{j=1}^3 \cos(\varphi_j-\varphi_{j-1}-\varphi/3)=H_C+H_J,
\end{align}
where $4E_C=\frac{1}{2}(2e)^2\gamma_1$ is the charging energy associated with the reciprocal capacitance $\gamma_1$ [see Eq.\ \eqref{gamma1}]. $\mathsf{M}$ is a dimensionless matrix obtained from the inverse capacitance matrix $\mathsf{C}^{-1}$ by rescaling and is defined as $(\mathsf{M})_{ij}=(1-\gamma)\delta_{ij} + \gamma$,
with $\gamma=\gamma_2/\gamma_1=E_C'/E_C$. 

Since $N$ has a discrete spectrum (comprised of all integers $\ZZ$), it is clear that the  offset-charge space spanned by $(n_{g1},n_{g2},n_{g3})$ is divided into regions of constant ground state charge number $N_0$. At the boundaries of these regions, $N_0$ must jump discontinuously. To understand the boundaries between such regions, we make an important observation which is not limited to the charging regime, but holds for \emph{arbitrary} $E_J/E_C$ ratio, and is also independent of all remaining model parameters: Any shift of the offset charges by integer amounts,
\be\label{ng1}
\vc{n}_g \, \to \, \vc{n}_g + (z_1,z_2,z_3) \qquad (z_i\in\ZZ)
\ee
leaves the spectrum of $H$ invariant and shifts $N_0$ according to 
\be\label{no-shift}
N_0\to N_0+\sum_i z_i.
\ee
 Further, at zero offset charge $\vc{n}_g=\vc{0}$, particle-hole symmetry is intact and dictates $N_0=0$. From Eq.\ \eqref{no-shift} one thus immediately knows that the ground state charge number obeys
\be
N_0(z_1,z_2,z_3) = z_1 + z_2 +z_3 \qquad  (z_i\in\ZZ).
\ee
Equation \eqref{ng1}, in fact, allows one to restrict the entire discussion to the domain $\vc{n}_{gj}\in[-1/2,1/2)$.
Symmetry also dictates that, assuming the simplest case of a direct transition from $N_0=0$ at $\vc{n}_g=\vc{0}$ to $N_0=\pm1$ at $\vc{n}_g=\pm \vc{e}_j$, the transition must occur at the midpoints. In other words, the points $\pm(1/2,0,0)$, $\pm(0,1/2,0)$ and $\pm(0,0,1/2)$ must lie on the boundaries separating $N_0=0$ from $N_0=\pm1$. Analogous arguments apply for the transition to $N_0=\pm1$ at $\vc{n}_g=\pm \vc{e}_1+\vc{e}_2-\vc{e}_3$ etc.\ along six out of the eight space diagonals, which puts the corresponding six corners of the unit cube on the boundaries. This sets the overall structure of $N_0$. The detailed form of the full boundary, however, depends on details such as the $E_J/E_C$ ratio. In the charge limit ($E_J\alt E_C$), $N_0$ can be constructed analytically and it is instructive to do so and to discuss how $N_0$ is modified for increased Josephson tunneling. 

In the charging regime, it is primarily the charging contribution $H_C$ which determines the boundaries between $N_0$ regions. To leading order, we  hence neglect Josephson tunneling ($H_J$) completely, and the problem becomes similar to the question of charge stability in a triple quantum dot \cite{rogge_three_2009}. The eigenstates of $H_C$  are charge eigenstates  $\ket{\vc{n}}$ with $\vc{n}\in\ZZ^3$ and corresponding eigenenergies $E_{\vc{n}}(\vc{n}_g)$. The boundary between the $N_0=0$ region centered at $\vc{n}_g=\vc{0}$ and the adjacent $N_0=\pm1$ regions reached via the planar diagonals are obtained by requiring that the respective energies match,
\be
E_{\vc{0}}(\vc{n}_g) = E_{\pm\vc{e}_j}(\vc{n}_g).
\ee
This yields six equations of the form
\be
0=1\mp 2(1-\gamma)n_{gj} \pm 2 \gamma \sum_k n_{gk} \quad (j=1,2,3),
\ee
which define planes in the offset charge space. Consistent truncation of the planes to the region where $N_0=0\to\pm1$ can occur, yields the full charge boundaries, see Fig. \ref{fig:ntot}. Note that in the charge regime, the $N_0$ boundaries do not depend on the magnetic flux.

The presence of Josephson tunneling will generally modify the shape of these boundaries, but leave the properties derived from general symmetry arguments intact. We expect $H_J$ to introduce flux-dependence and to smoothen the sharp-edge boundaries [see, e.g., Fig.\ \ref{fig:ntot}(a) and (b)], as it hybridizes the states $\ket{\vc{e}_j}$ for $j=1,2,3$ and thus turns crossings into avoided crossings.


\begin{thebibliography}{71}
\expandafter\ifx\csname natexlab\endcsname\relax\def\natexlab#1{#1}\fi
\expandafter\ifx\csname bibnamefont\endcsname\relax
  \def\bibnamefont#1{#1}\fi
\expandafter\ifx\csname bibfnamefont\endcsname\relax
  \def\bibfnamefont#1{#1}\fi
\expandafter\ifx\csname citenamefont\endcsname\relax
  \def\citenamefont#1{#1}\fi
\expandafter\ifx\csname url\endcsname\relax
  \def\url#1{\texttt{#1}}\fi
\expandafter\ifx\csname urlprefix\endcsname\relax\def\urlprefix{URL }\fi
\providecommand{\bibinfo}[2]{#2}
\providecommand{\eprint}[2][]{\url{#2}}

\bibitem[{\citenamefont{Greentree et~al.}(2006)\citenamefont{Greentree, Tahan,
  Cole, and Hollenberg}}]{greentree_quantum_2006}
\bibinfo{author}{\bibfnamefont{A.~D.} \bibnamefont{Greentree}},
  \bibinfo{author}{\bibfnamefont{C.}~\bibnamefont{Tahan}},
  \bibinfo{author}{\bibfnamefont{J.~H.} \bibnamefont{Cole}}, \bibnamefont{and}
  \bibinfo{author}{\bibfnamefont{L.~C.~L.} \bibnamefont{Hollenberg}},
  \bibinfo{journal}{Nat. Phys.} \textbf{\bibinfo{volume}{2}},
  \bibinfo{pages}{856} (\bibinfo{year}{2006}).

\bibitem[{\citenamefont{Hartmann et~al.}(2006)\citenamefont{Hartmann, Brandão,
  and Plenio}}]{hartmann_strongly_2006}
\bibinfo{author}{\bibfnamefont{M.~J.} \bibnamefont{Hartmann}},
  \bibinfo{author}{\bibfnamefont{F.~G. S.~L.} \bibnamefont{Brandão}},
  \bibnamefont{and} \bibinfo{author}{\bibfnamefont{M.~B.}
  \bibnamefont{Plenio}}, \bibinfo{journal}{Nat. Phys.}
  \textbf{\bibinfo{volume}{2}}, \bibinfo{pages}{849}
  (\bibinfo{year}{2006}).
  
  \bibitem[{\citenamefont{Angelakis et~al.}(2007)\citenamefont{Angelakis, Santos,
  and Bose}}]{angelakis_photon-blockade-induced_2007}
\bibinfo{author}{\bibfnamefont{D.~G.} \bibnamefont{Angelakis}},
  \bibinfo{author}{\bibfnamefont{M.~F.} \bibnamefont{Santos}},
  \bibnamefont{and} \bibinfo{author}{\bibfnamefont{S.}~\bibnamefont{Bose}},
  \bibinfo{journal}{Phys. Rev. A} \textbf{\bibinfo{volume}{76}},
  \bibinfo{pages}{031805} (\bibinfo{year}{2007}).

\bibitem[{\citenamefont{Buluta and Nori}(2009)}]{buluta_quantum_2009}
\bibinfo{author}{\bibfnamefont{I.}~\bibnamefont{Buluta}} \bibnamefont{and}
  \bibinfo{author}{\bibfnamefont{F.}~\bibnamefont{Nori}},
  \bibinfo{journal}{Science} \textbf{\bibinfo{volume}{326}},
  \bibinfo{pages}{108} (\bibinfo{year}{2009}).

\bibitem[{\citenamefont{Hartmann et~al.}(2008)\citenamefont{Hartmann, Brandão,
  and Plenio}}]{hartmann_quantum_2008}
\bibinfo{author}{\bibfnamefont{M.~J.} \bibnamefont{Hartmann}},
  \bibinfo{author}{\bibfnamefont{F.~G. S.~L.} \bibnamefont{Brandão}},
  \bibnamefont{and} \bibinfo{author}{\bibfnamefont{M.~B.}
  \bibnamefont{Plenio}}, \bibinfo{journal}{Laser Photonics Rev.}
  \textbf{\bibinfo{volume}{2}}, \bibinfo{pages}{527}
  (\bibinfo{year}{2008}).

\bibitem[{\citenamefont{{{Le} Hur}}(2009)}]{le_hur_quantum_2009}
\bibinfo{author}{\bibfnamefont{K.}~\bibnamefont{{{Le} Hur}}},
  \bibinfo{journal}{arXiv:0909.4822}  (\bibinfo{year}{2009}).

\bibitem[{\citenamefont{Tomadin and Fazio}(2010)}]{tomadin_many-body_2010}
\bibinfo{author}{\bibfnamefont{A.}~\bibnamefont{Tomadin}} \bibnamefont{and}
  \bibinfo{author}{\bibfnamefont{R.}~\bibnamefont{Fazio}}, \bibinfo{journal}{J.
  Opt. Soc. Am. B: Opt. Phys.} \textbf{\bibinfo{volume}{27}},
  \bibinfo{pages}{A130} (\bibinfo{year}{2010}).

\bibitem[{\citenamefont{Lewenstein et~al.}(2007)\citenamefont{Lewenstein,
  Sanpera, Ahufinger, Damski, Sen, and Sen}}]{lewenstein_ultracold_2007}
\bibinfo{author}{\bibfnamefont{M.}~\bibnamefont{Lewenstein}},
  \bibinfo{author}{\bibfnamefont{A.}~\bibnamefont{Sanpera}},
  \bibinfo{author}{\bibfnamefont{V.}~\bibnamefont{Ahufinger}},
  \bibinfo{author}{\bibfnamefont{B.}~\bibnamefont{Damski}},
  \bibinfo{author}{\bibfnamefont{A.}~\bibnamefont{Sen}}, \bibnamefont{and}
  \bibinfo{author}{\bibfnamefont{U.}~\bibnamefont{Sen}},
  \bibinfo{journal}{Adv. Phys.} \textbf{\bibinfo{volume}{56}},
  \bibinfo{pages}{243} (\bibinfo{year}{2007}).

\bibitem[{\citenamefont{Bloch et~al.}(2008)\citenamefont{Bloch, Dalibard, and
  Zwerger}}]{bloch_many-body_2008}
\bibinfo{author}{\bibfnamefont{I.}~\bibnamefont{Bloch}},
  \bibinfo{author}{\bibfnamefont{J.}~\bibnamefont{Dalibard}}, \bibnamefont{and}
  \bibinfo{author}{\bibfnamefont{W.}~\bibnamefont{Zwerger}},
  \bibinfo{journal}{Rev. Mod. Phys.} \textbf{\bibinfo{volume}{80}},
  \bibinfo{pages}{885} (\bibinfo{year}{2008}).

\bibitem[{\citenamefont{Rossini and
  Fazio}(2007)}]{rossini_mott-insulating_2007}
\bibinfo{author}{\bibfnamefont{D.}~\bibnamefont{Rossini}} \bibnamefont{and}
  \bibinfo{author}{\bibfnamefont{R.}~\bibnamefont{Fazio}},
  \bibinfo{journal}{Phys. Rev. Lett.} \textbf{\bibinfo{volume}{99}},
  \bibinfo{pages}{186401} (\bibinfo{year}{2007}).

\bibitem[{\citenamefont{Rossini et~al.}(2008)\citenamefont{Rossini, Fazio, and
  Santoro}}]{rossini_photon_2008}
\bibinfo{author}{\bibfnamefont{D.}~\bibnamefont{Rossini}},
  \bibinfo{author}{\bibfnamefont{R.}~\bibnamefont{Fazio}}, \bibnamefont{and}
  \bibinfo{author}{\bibfnamefont{G.}~\bibnamefont{Santoro}},
  \bibinfo{journal}{Europhys. Lett.} \textbf{\bibinfo{volume}{83}},
  \bibinfo{pages}{47011} (\bibinfo{year}{2008}).

\bibitem[{\citenamefont{Na et~al.}(2008)\citenamefont{Na, Utsunomiya, Tian, and
  Yamamoto}}]{na_strongly_2008-1}
\bibinfo{author}{\bibfnamefont{N.}~\bibnamefont{Na}},
  \bibinfo{author}{\bibfnamefont{S.}~\bibnamefont{Utsunomiya}},
  \bibinfo{author}{\bibfnamefont{L.}~\bibnamefont{Tian}}, \bibnamefont{and}
  \bibinfo{author}{\bibfnamefont{Y.}~\bibnamefont{Yamamoto}},
  \bibinfo{journal}{Phys. Rev. A} \textbf{\bibinfo{volume}{77}},
  \bibinfo{pages}{031803} (\bibinfo{year}{2008}).

\bibitem[{\citenamefont{Aichhorn et~al.}(2008)\citenamefont{Aichhorn,
  Hohenadler, Tahan, and Littlewood}}]{aichhorn_quantum_2008}
\bibinfo{author}{\bibfnamefont{M.}~\bibnamefont{Aichhorn}},
  \bibinfo{author}{\bibfnamefont{M.}~\bibnamefont{Hohenadler}},
  \bibinfo{author}{\bibfnamefont{C.}~\bibnamefont{Tahan}}, \bibnamefont{and}
  \bibinfo{author}{\bibfnamefont{P.~B.} \bibnamefont{Littlewood}},
  \bibinfo{journal}{Phys. Rev. Lett.} \textbf{\bibinfo{volume}{100}},
  \bibinfo{pages}{216401} (\bibinfo{year}{2008}).

\bibitem[{\citenamefont{Zhao et~al.}(2008)\citenamefont{Zhao, Sandvik, and
  Ueda}}]{zhao_insulator_2008}
\bibinfo{author}{\bibfnamefont{J.}~\bibnamefont{Zhao}},
  \bibinfo{author}{\bibfnamefont{A.~W.} \bibnamefont{Sandvik}},
  \bibnamefont{and} \bibinfo{author}{\bibfnamefont{K.}~\bibnamefont{Ueda}},
  \bibinfo{journal}{arXiv:0806.3603}  (\bibinfo{year}{2008}).

\bibitem[{\citenamefont{Koch and {{Le}
  Hur}}(2009)}]{koch_superfluidmott-insulator_2009}
\bibinfo{author}{\bibfnamefont{J.}~\bibnamefont{Koch}} \bibnamefont{and}
  \bibinfo{author}{\bibfnamefont{K.}~\bibnamefont{{{Le} Hur}}},
  \bibinfo{journal}{Phys. Rev. A} \textbf{\bibinfo{volume}{80}},
  \bibinfo{pages}{023811} (\bibinfo{year}{2009}).

\bibitem[{\citenamefont{Schmidt and Blatter}(2009)}]{schmidt_strong_2009-1}
\bibinfo{author}{\bibfnamefont{S.}~\bibnamefont{Schmidt}} \bibnamefont{and}
  \bibinfo{author}{\bibfnamefont{G.}~\bibnamefont{Blatter}},
  \bibinfo{journal}{Phys. Rev. Lett.} \textbf{\bibinfo{volume}{103}},
  \bibinfo{pages}{086403} (\bibinfo{year}{2009}).

\bibitem[{\citenamefont{Schmidt and
  Blatter}(2010)}]{schmidt_excitations_2010-1}
\bibinfo{author}{\bibfnamefont{S.}~\bibnamefont{Schmidt}} \bibnamefont{and}
  \bibinfo{author}{\bibfnamefont{G.}~\bibnamefont{Blatter}},
  \bibinfo{journal}{Phys. Rev. Lett.} \textbf{\bibinfo{volume}{104}},
  \bibinfo{pages}{216402} (\bibinfo{year}{2010}).


\bibitem[{\citenamefont{Fisher et~al.}(1989)\citenamefont{Fisher, Weichman,
  Grinstein, and Fisher}}]{fisher_boson_1989}
\bibinfo{author}{\bibfnamefont{M.~P.~A.} \bibnamefont{Fisher}},
  \bibinfo{author}{\bibfnamefont{P.~B.} \bibnamefont{Weichman}},
  \bibinfo{author}{\bibfnamefont{G.}~\bibnamefont{Grinstein}},
  \bibnamefont{and} \bibinfo{author}{\bibfnamefont{D.~S.}
  \bibnamefont{Fisher}}, \bibinfo{journal}{Phys. Rev. B}
  \textbf{\bibinfo{volume}{40}}, \bibinfo{pages}{546} (\bibinfo{year}{1989}).

\bibitem[{\citenamefont{Sachdev}(2000)}]{sachdev_quantum_2000}
\bibinfo{author}{\bibfnamefont{S.}~\bibnamefont{Sachdev}},
  \emph{\bibinfo{title}{Quantum Phase Transitions}}
  (\bibinfo{publisher}{Cambridge University Press}, \bibinfo{year}{2000}).

\bibitem[{\citenamefont{Bruder et~al.}(2005)\citenamefont{Bruder, Fazio, and
  Schön}}]{bruder_bose-hubbard_2005}
\bibinfo{author}{\bibfnamefont{C.}~\bibnamefont{Bruder}},
  \bibinfo{author}{\bibfnamefont{R.}~\bibnamefont{Fazio}}, \bibnamefont{and}
  \bibinfo{author}{\bibfnamefont{G.}~\bibnamefont{Schön}},
  \bibinfo{journal}{Ann. der Physik} \textbf{\bibinfo{volume}{14}},
  \bibinfo{pages}{566} (\bibinfo{year}{2005}).

\bibitem[{\citenamefont{Tomadin et~al.}(2009)\citenamefont{Tomadin,
  Giovannetti, Fazio, Gerace, Carusotto, Tureci, and
  Imamoglu}}]{tomadin_non-equilibrium_2009}
\bibinfo{author}{\bibfnamefont{A.}~\bibnamefont{Tomadin}},
  \bibinfo{author}{\bibfnamefont{V.}~\bibnamefont{Giovannetti}},
  \bibinfo{author}{\bibfnamefont{R.}~\bibnamefont{Fazio}},
  \bibinfo{author}{\bibfnamefont{D.}~\bibnamefont{Gerace}},
  \bibinfo{author}{\bibfnamefont{I.}~\bibnamefont{Carusotto}},
  \bibinfo{author}{\bibfnamefont{H.~E.} \bibnamefont{Tureci}},
  \bibnamefont{and} \bibinfo{author}{\bibfnamefont{A.}~\bibnamefont{Imamoglu}},
  \bibinfo{journal}{arXiv:0904.4437}  (\bibinfo{year}{2009}).

\bibitem[{\citenamefont{Carusotto et~al.}(2009)\citenamefont{Carusotto, Gerace,
  Tureci, Liberato, Ciuti, and Imamoglu}}]{carusotto_fermionized_2009}
\bibinfo{author}{\bibfnamefont{I.}~\bibnamefont{Carusotto}},
  \bibinfo{author}{\bibfnamefont{D.}~\bibnamefont{Gerace}},
  \bibinfo{author}{\bibfnamefont{H.~E.} \bibnamefont{Tureci}},
  \bibinfo{author}{\bibfnamefont{S.~D.} \bibnamefont{Liberato}},
  \bibinfo{author}{\bibfnamefont{C.}~\bibnamefont{Ciuti}}, \bibnamefont{and}
  \bibinfo{author}{\bibfnamefont{A.}~\bibnamefont{Imamoglu}},
  \bibinfo{journal}{Phys. Rev. Lett.} \textbf{\bibinfo{volume}{103}},
  \bibinfo{pages}{033601} (\bibinfo{year}{2009}).

\bibitem[{\citenamefont{Kiffner and
  Hartmann}(2010)}]{kiffner_dissipation-induced_2010}
\bibinfo{author}{\bibfnamefont{M.}~\bibnamefont{Kiffner}} \bibnamefont{and}
  \bibinfo{author}{\bibfnamefont{M.~J.} \bibnamefont{Hartmann}},
  \bibinfo{journal}{Phys. Rev. A} \textbf{\bibinfo{volume}{81}},
  \bibinfo{pages}{021806} (\bibinfo{year}{2010}).

\bibitem[{\citenamefont{Hartmann}(2010)}]{hartmann_polariton_2010}
\bibinfo{author}{\bibfnamefont{M.~J.} \bibnamefont{Hartmann}},
  \bibinfo{journal}{Phys. Rev. Lett.} \textbf{\bibinfo{volume}{104}},
  \bibinfo{pages}{113601} (\bibinfo{year}{2010}).

\bibitem[{\citenamefont{Tsui et~al.}(1982)\citenamefont{Tsui, Stormer, and
  Gossard}}]{tsui_two-dimensional_1982}
\bibinfo{author}{\bibfnamefont{D.~C.} \bibnamefont{Tsui}},
  \bibinfo{author}{\bibfnamefont{H.~L.} \bibnamefont{Stormer}},
  \bibnamefont{and} \bibinfo{author}{\bibfnamefont{A.~C.}
  \bibnamefont{Gossard}}, \bibinfo{journal}{Phys. Rev. Lett.}
  \textbf{\bibinfo{volume}{48}}, \bibinfo{pages}{1559} (\bibinfo{year}{1982}).

\bibitem[{\citenamefont{Laughlin}(1983)}]{laughlin_anomalous_1983-1}
\bibinfo{author}{\bibfnamefont{R.~B.} \bibnamefont{Laughlin}},
  \bibinfo{journal}{Phys. Rev. Lett.} \textbf{\bibinfo{volume}{50}},
  \bibinfo{pages}{1395} (\bibinfo{year}{1983}).

\bibitem[{\citenamefont{Jaksch and Zoller}(2003)}]{jaksch_creation_2003}
\bibinfo{author}{\bibfnamefont{D.}~\bibnamefont{Jaksch}} \bibnamefont{and}
  \bibinfo{author}{\bibfnamefont{P.}~\bibnamefont{Zoller}},
  \bibinfo{journal}{New J. Phys.} \textbf{\bibinfo{volume}{5}},
  \bibinfo{pages}{56} (\bibinfo{year}{2003}).

\bibitem[{\citenamefont{Paredes et~al.}(2003)\citenamefont{Paredes, Zoller, and
  Cirac}}]{paredes_fractional_2003}
\bibinfo{author}{\bibfnamefont{B.}~\bibnamefont{Paredes}},
  \bibinfo{author}{\bibfnamefont{P.}~\bibnamefont{Zoller}}, \bibnamefont{and}
  \bibinfo{author}{\bibfnamefont{J.~I.} \bibnamefont{Cirac}},
  \bibinfo{journal}{Solid State Commun.} \textbf{\bibinfo{volume}{127}},
  \bibinfo{pages}{155} (\bibinfo{year}{2003}).

\bibitem[{\citenamefont{Sørensen et~al.}(2005)\citenamefont{Sørensen, Demler,
  and Lukin}}]{srensen_fractional_2005}
\bibinfo{author}{\bibfnamefont{A.~S.} \bibnamefont{Sørensen}},
  \bibinfo{author}{\bibfnamefont{E.}~\bibnamefont{Demler}}, \bibnamefont{and}
  \bibinfo{author}{\bibfnamefont{M.~D.} \bibnamefont{Lukin}},
  \bibinfo{journal}{Phys. Rev. Lett.} \textbf{\bibinfo{volume}{94}},
  \bibinfo{pages}{086803} (\bibinfo{year}{2005}).

\bibitem[{\citenamefont{Lin et~al.}(2009)\citenamefont{Lin, Compton,
  {Jimenez-Garcia}, Porto, and Spielman}}]{lin_synthetic_2009}
\bibinfo{author}{\bibfnamefont{Y.}~\bibnamefont{Lin}},
  \bibinfo{author}{\bibfnamefont{R.~L.} \bibnamefont{Compton}},
  \bibinfo{author}{\bibfnamefont{K.}~\bibnamefont{{Jimenez-Garcia}}},
  \bibinfo{author}{\bibfnamefont{J.~V.} \bibnamefont{Porto}}, \bibnamefont{and}
  \bibinfo{author}{\bibfnamefont{I.~B.} \bibnamefont{Spielman}},
  \bibinfo{journal}{Nature (London)} \textbf{\bibinfo{volume}{462}},
  \bibinfo{pages}{628} (\bibinfo{year}{2009}).

\bibitem[{\citenamefont{Cho et~al.}(2008)\citenamefont{Cho, Angelakis, and
  Bose}}]{cho_fractional_2008}
\bibinfo{author}{\bibfnamefont{J.}~\bibnamefont{Cho}},
  \bibinfo{author}{\bibfnamefont{D.~G.} \bibnamefont{Angelakis}},
  \bibnamefont{and} \bibinfo{author}{\bibfnamefont{S.}~\bibnamefont{Bose}},
  \bibinfo{journal}{Phys. Rev. Lett.} \textbf{\bibinfo{volume}{101}},
  \bibinfo{pages}{246809} (\bibinfo{year}{2008}).

\bibitem[{\citenamefont{Haldane and Raghu}(2008)}]{haldane_possible_2008}
\bibinfo{author}{\bibfnamefont{F.~D.~M.} \bibnamefont{Haldane}}
  \bibnamefont{and} \bibinfo{author}{\bibfnamefont{S.}~\bibnamefont{Raghu}},
  \bibinfo{journal}{Phys. Rev. Lett.} \textbf{\bibinfo{volume}{100}},
  \bibinfo{pages}{013904} (\bibinfo{year}{2008}).

\bibitem[{\citenamefont{Raghu and Haldane}(2008)}]{raghu_analogs_2008}
\bibinfo{author}{\bibfnamefont{S.}~\bibnamefont{Raghu}} \bibnamefont{and}
  \bibinfo{author}{\bibfnamefont{F.~D.~M.} \bibnamefont{Haldane}},
  \bibinfo{journal}{Phys. Rev. A} \textbf{\bibinfo{volume}{78}},
  \bibinfo{pages}{033834} (\bibinfo{year}{2008}).
  
  \bibitem[{\citenamefont{Wang et~al.}(2009)}]{zwang}
\bibinfo{author}{\bibfnamefont{Z.}~\bibnamefont{Wang}},
\bibinfo{author}{\bibfnamefont{Y.}~\bibnamefont{Chong}},
\bibinfo{author}{\bibfnamefont{Y.~D.}~\bibnamefont{Joannopoulos}}, \bibnamefont{and}
  \bibinfo{author}{\bibfnamefont{M.} \bibnamefont{Solja\v{c}i\'c}},
  \bibinfo{journal}{Nature (London)} \textbf{\bibinfo{volume}{461}},
  \bibinfo{pages}{772} (\bibinfo{year}{2009}).

\bibitem[{\citenamefont{Blais et~al.}(2004)\citenamefont{Blais, Huang,
  Wallraff, Girvin, and Schoelkopf}}]{blais_cavity_2004}
\bibinfo{author}{\bibfnamefont{A.}~\bibnamefont{Blais}},
  \bibinfo{author}{\bibfnamefont{R.}~\bibnamefont{Huang}},
  \bibinfo{author}{\bibfnamefont{A.}~\bibnamefont{Wallraff}},
  \bibinfo{author}{\bibfnamefont{S.~M.} \bibnamefont{Girvin}},
  \bibnamefont{and} \bibinfo{author}{\bibfnamefont{R.~J.}
  \bibnamefont{Schoelkopf}}, \bibinfo{journal}{Phys. Rev. A}
  \textbf{\bibinfo{volume}{69}}, \bibinfo{pages}{062320}
  (\bibinfo{year}{2004}).

\bibitem[{\citenamefont{Wallraff et~al.}(2004)\citenamefont{Wallraff, Schuster,
  Blais, Frunzio, Majer, Kumar, Girvin, and Schoelkopf}}]{wallraff_strong_2004}
\bibinfo{author}{\bibfnamefont{A.}~\bibnamefont{Wallraff}},
  \bibinfo{author}{\bibfnamefont{D.~I.} \bibnamefont{Schuster}},
  \bibinfo{author}{\bibfnamefont{A.}~\bibnamefont{Blais}},
  \bibinfo{author}{\bibfnamefont{L.}~\bibnamefont{Frunzio}},
  \bibinfo{author}{\bibfnamefont{R.~H.~J.} \bibnamefont{Majer}},
  \bibinfo{author}{\bibfnamefont{S.}~\bibnamefont{Kumar}},
  \bibinfo{author}{\bibfnamefont{S.~M.} \bibnamefont{Girvin}},
  \bibnamefont{and} \bibinfo{author}{\bibfnamefont{R.~J.}
  \bibnamefont{Schoelkopf}}, \bibinfo{journal}{Nature {(London)}}
  \textbf{\bibinfo{volume}{431}}, \bibinfo{pages}{162} (\bibinfo{year}{2004}).

\bibitem[{\citenamefont{Schoelkopf and Girvin}(2008)}]{schoelkopf_wiring_2008}
\bibinfo{author}{\bibfnamefont{R.~J.} \bibnamefont{Schoelkopf}}
  \bibnamefont{and} \bibinfo{author}{\bibfnamefont{S.~M.}
  \bibnamefont{Girvin}}, \bibinfo{journal}{Nature {(London)}}
  \textbf{\bibinfo{volume}{451}}, \bibinfo{pages}{664} (\bibinfo{year}{2008}).

\bibitem[{\citenamefont{Schrieffer and Wolff}(1966)}]{schrieffer_relation_1966}
\bibinfo{author}{\bibfnamefont{J.~R.} \bibnamefont{Schrieffer}}
  \bibnamefont{and} \bibinfo{author}{\bibfnamefont{P.~A.} \bibnamefont{Wolff}},
  \bibinfo{journal}{Phys. Rev.} \textbf{\bibinfo{volume}{149}},
  \bibinfo{pages}{491} (\bibinfo{year}{1966}).

\bibitem[{\citenamefont{{Cohen-Tannoudji}
  et~al.}(1998)\citenamefont{{Cohen-Tannoudji}, {Dupont-Roc}, and
  Grynberg}}]{cohen-tannoudji_atom-photon_1998}
\bibinfo{author}{\bibfnamefont{C.}~\bibnamefont{{Cohen-Tannoudji}}},
  \bibinfo{author}{\bibfnamefont{J.}~\bibnamefont{{Dupont-Roc}}},
  \bibnamefont{and} \bibinfo{author}{\bibfnamefont{G.}~\bibnamefont{Grynberg}},
  \emph{\bibinfo{title}{{Atom-Photon} Interactions: Basic Processes and
  Applications}} (\bibinfo{publisher}{{Wiley-VCH}}, \bibinfo{year}{1998}), Chap.\ $\text{B}_\text{I}$.

\bibitem[{\citenamefont{Pozar}(2004)}]{pozar_microwave_2004}
\bibinfo{author}{\bibfnamefont{D.~M.} \bibnamefont{Pozar}},
  \emph{\bibinfo{title}{Microwave Engineering}} (\bibinfo{publisher}{Wiley},
  \bibinfo{year}{2004}), \bibinfo{edition}{3rd} ed.

\bibitem[{\citenamefont{Walls and Milburn}(1995)}]{walls_quantum_1995}
\bibinfo{author}{\bibfnamefont{D.}~\bibnamefont{Walls}} \bibnamefont{and}
  \bibinfo{author}{\bibfnamefont{G.}~\bibnamefont{Milburn}},
  \emph{\bibinfo{title}{Quantum Optics}} (\bibinfo{publisher}{Springer},
  \bibinfo{year}{1995}), \bibinfo{edition}{1st} ed.

\bibitem[{\citenamefont{Clerk et~al.}(2008)\citenamefont{Clerk, Devoret,
  Girvin, Marquardt, and Schoelkopf}}]{clerk_introduction_2008}
\bibinfo{author}{\bibfnamefont{A.~A.} \bibnamefont{Clerk}},
  \bibinfo{author}{\bibfnamefont{M.~H.} \bibnamefont{Devoret}},
  \bibinfo{author}{\bibfnamefont{S.~M.} \bibnamefont{Girvin}},
  \bibinfo{author}{\bibfnamefont{F.}~\bibnamefont{Marquardt}},
  \bibnamefont{and} \bibinfo{author}{\bibfnamefont{R.~J.}
  \bibnamefont{Schoelkopf}},   \bibinfo{journal}{Rev. Mod. Phys.} \textbf{\bibinfo{volume}{82}},
  \bibinfo{pages}{1155} (\bibinfo{year}{2010}) and \bibinfo{journal}{arXiv:0810.4729}
  (\bibinfo{year}{2008}).

\bibitem[{\citenamefont{Mekata}(2003)}]{mekata_kagome:story_2003}
\bibinfo{author}{\bibfnamefont{M.}~\bibnamefont{Mekata}},
  \bibinfo{journal}{Physics Today} \textbf{\bibinfo{volume}{56}},
  \bibinfo{pages}{12} (\bibinfo{year}{2003}).

\bibitem[{\citenamefont{Syôzi}(1951)}]{syzi_statistics_1951}
\bibinfo{author}{\bibfnamefont{I.}~\bibnamefont{Syôzi}},
  \bibinfo{journal}{Prog. Theor. Phys.} \textbf{\bibinfo{volume}{6}},
  \bibinfo{pages}{306} (\bibinfo{year}{1951}).

\bibitem[{\citenamefont{Kanô and
  Naya}(1953)}]{kan_antiferromagnetism.kagom_1953}
\bibinfo{author}{\bibfnamefont{K.}~\bibnamefont{Kanô}} \bibnamefont{and}
  \bibinfo{author}{\bibfnamefont{S.}~\bibnamefont{Naya}},
  \bibinfo{journal}{Prog. Theor. Phys.} \textbf{\bibinfo{volume}{10}},
  \bibinfo{pages}{158} (\bibinfo{year}{1953}).

\bibitem[{\citenamefont{Wolf and Schotte}(1988)}]{wolf_ising_1988}
\bibinfo{author}{\bibfnamefont{M.}~\bibnamefont{Wolf}} \bibnamefont{and}
  \bibinfo{author}{\bibfnamefont{K.~D.} \bibnamefont{Schotte}},
  \bibinfo{journal}{J. Phys. A: Math. Gen.} \textbf{\bibinfo{volume}{21}}, \bibinfo{pages}{2195} (\bibinfo{year}{1988}).

\bibitem[{\citenamefont{Lecheminant et~al.}(1997)\citenamefont{Lecheminant,
  Bernu, Lhuillier, Pierre, and Sindzingre}}]{lecheminant_order_1997}
\bibinfo{author}{\bibfnamefont{P.}~\bibnamefont{Lecheminant}},
  \bibinfo{author}{\bibfnamefont{B.}~\bibnamefont{Bernu}},
  \bibinfo{author}{\bibfnamefont{C.}~\bibnamefont{Lhuillier}},
  \bibinfo{author}{\bibfnamefont{L.}~\bibnamefont{Pierre}}, \bibnamefont{and}
  \bibinfo{author}{\bibfnamefont{P.}~\bibnamefont{Sindzingre}},
  \bibinfo{journal}{Phys. Rev. B} \textbf{\bibinfo{volume}{56}},
  \bibinfo{pages}{2521} (\bibinfo{year}{1997}).

\bibitem[{\citenamefont{Waldtmann et~al.}(1998)\citenamefont{Waldtmann, Everts,
  Bernu, Lhuillier, Sindzingre, Lecheminant, and
  Pierre}}]{waldtmann_first_1998}
\bibinfo{author}{\bibfnamefont{C.}~\bibnamefont{Waldtmann}},
  \bibinfo{author}{\bibfnamefont{H.}~\bibnamefont{Everts}},
  \bibinfo{author}{\bibfnamefont{B.}~\bibnamefont{Bernu}},
  \bibinfo{author}{\bibfnamefont{C.}~\bibnamefont{Lhuillier}},
  \bibinfo{author}{\bibfnamefont{P.}~\bibnamefont{Sindzingre}},
  \bibinfo{author}{\bibfnamefont{P.}~\bibnamefont{Lecheminant}},
  \bibnamefont{and} \bibinfo{author}{\bibfnamefont{L.}~\bibnamefont{Pierre}},
  \bibinfo{journal}{Eur. Phys. J. B}
  \textbf{\bibinfo{volume}{2}}, \bibinfo{pages}{501}
  (\bibinfo{year}{1998}).

\bibitem[{\citenamefont{Balents et~al.}(2002)\citenamefont{Balents, Fisher, and
  Girvin}}]{balents_fractionalization_2002}
\bibinfo{author}{\bibfnamefont{L.}~\bibnamefont{Balents}},
  \bibinfo{author}{\bibfnamefont{M.~P.~A.} \bibnamefont{Fisher}},
  \bibnamefont{and} \bibinfo{author}{\bibfnamefont{S.~M.}
  \bibnamefont{Girvin}}, \bibinfo{journal}{Phys. Rev. B}
  \textbf{\bibinfo{volume}{65}}, \bibinfo{pages}{224412}
  (\bibinfo{year}{2002}).

\bibitem[{\citenamefont{Mielke}(1991)}]{mielke__1991}
\bibinfo{author}{\bibfnamefont{A.}~\bibnamefont{Mielke}}, \bibinfo{journal}{J.
  Phys. A: Math. Gen.} \textbf{\bibinfo{volume}{24}}, \bibinfo{pages}{L73}
  (\bibinfo{year}{1991}).

\bibitem[{\citenamefont{Mielke}(1992)}]{mielke__1992}
\bibinfo{author}{\bibfnamefont{A.}~\bibnamefont{Mielke}}, \bibinfo{journal}{J.
  Phys. A: Math. Gen.} \textbf{\bibinfo{volume}{25}}, \bibinfo{pages}{4335}
  (\bibinfo{year}{1992}).

\bibitem[{\citenamefont{Mielke and Tasaki}(1993)}]{mielke_ferromagnetism_1993}
\bibinfo{author}{\bibfnamefont{A.}~\bibnamefont{Mielke}} \bibnamefont{and}
  \bibinfo{author}{\bibfnamefont{H.}~\bibnamefont{Tasaki}},
  \bibinfo{journal}{Commun. Math. Phys.}
  \textbf{\bibinfo{volume}{158}}, \bibinfo{pages}{341} (\bibinfo{year}{1993}).

\bibitem[{\citenamefont{Santos et~al.}(2004)\citenamefont{Santos, Baranov,
  Cirac, Everts, Fehrmann, and Lewenstein}}]{santos_atomic_2004}
\bibinfo{author}{\bibfnamefont{L.}~\bibnamefont{Santos}},
  \bibinfo{author}{\bibfnamefont{M.~A.} \bibnamefont{Baranov}},
  \bibinfo{author}{\bibfnamefont{J.~I.} \bibnamefont{Cirac}},
  \bibinfo{author}{\bibfnamefont{H.}~\bibnamefont{Everts}},
  \bibinfo{author}{\bibfnamefont{H.}~\bibnamefont{Fehrmann}}, \bibnamefont{and}
  \bibinfo{author}{\bibfnamefont{M.}~\bibnamefont{Lewenstein}},
  \bibinfo{journal}{Phys. Rev. Lett.} \textbf{\bibinfo{volume}{93}},
  \bibinfo{pages}{030601} (\bibinfo{year}{2004}).

\bibitem[{\citenamefont{Isakov et~al.}(2006)\citenamefont{Isakov, Wessel,
  Melko, Sengupta, and Kim}}]{isakov_hard-core_2006}
\bibinfo{author}{\bibfnamefont{S.~V.} \bibnamefont{Isakov}},
  \bibinfo{author}{\bibfnamefont{S.}~\bibnamefont{Wessel}},
  \bibinfo{author}{\bibfnamefont{R.~G.} \bibnamefont{Melko}},
  \bibinfo{author}{\bibfnamefont{K.}~\bibnamefont{Sengupta}}, \bibnamefont{and}
  \bibinfo{author}{\bibfnamefont{Y.~B.} \bibnamefont{Kim}},
  \bibinfo{journal}{Phys. Rev. Lett.} \textbf{\bibinfo{volume}{97}},
  \bibinfo{pages}{147202} (\bibinfo{year}{2006}).

\bibitem[{\citenamefont{Guo and Franz}(2009)}]{guo_topological_2009}
\bibinfo{author}{\bibfnamefont{H.}~\bibnamefont{Guo}} \bibnamefont{and}
  \bibinfo{author}{\bibfnamefont{M.}~\bibnamefont{Franz}},
  \bibinfo{journal}{Phys. Rev. B} \textbf{\bibinfo{volume}{80}},
  \bibinfo{pages}{113102} (\bibinfo{year}{2009}).

\bibitem[{\citenamefont{Wen et~al.}(2010)\citenamefont{Wen, Rüegg, Wang, and
  Fiete}}]{wen_interaction-driven_2010}
\bibinfo{author}{\bibfnamefont{J.}~\bibnamefont{Wen}},
  \bibinfo{author}{\bibfnamefont{A.}~\bibnamefont{Rüegg}},
  \bibinfo{author}{\bibfnamefont{C.~C.~J.} \bibnamefont{Wang}},
  \bibnamefont{and} \bibinfo{author}{\bibfnamefont{G.~A.} \bibnamefont{Fiete}},
  \bibinfo{journal}{{arXiv:1005.4061}}  (\bibinfo{year}{2010}).

\bibitem[{\citenamefont{Nishino et~al.}(2003)\citenamefont{Nishino, Goda, and
  Kusakabe}}]{nishino_flat_2003}
\bibinfo{author}{\bibfnamefont{S.}~\bibnamefont{Nishino}},
  \bibinfo{author}{\bibfnamefont{M.}~\bibnamefont{Goda}}, \bibnamefont{and}
  \bibinfo{author}{\bibfnamefont{K.}~\bibnamefont{Kusakabe}},
  \bibinfo{journal}{J. Phys. Soc. Jpn.} \textbf{\bibinfo{volume}{72}},
  \bibinfo{pages}{2015} (\bibinfo{year}{2003}).

\bibitem[{\citenamefont{Green et~al.}(2010)\citenamefont{Green, Santos, and
  Chamon}}]{green_isolated_2010}
\bibinfo{author}{\bibfnamefont{D.}~\bibnamefont{Green}},
  \bibinfo{author}{\bibfnamefont{L.}~\bibnamefont{Santos}}, \bibnamefont{and}
  \bibinfo{author}{\bibfnamefont{C.}~\bibnamefont{Chamon}},
  \bibinfo{journal}{{arXiv:1004.0708}}  (\bibinfo{year}{2010}).

\bibitem[{\citenamefont{Devoret}(1997)}]{devoret_quantum_1997}
\bibinfo{author}{\bibfnamefont{M.~H.} \bibnamefont{Devoret}}, in
  \emph{\bibinfo{booktitle}{Quantum Fluctuations {(Les} Houches Session
  {LXIII)}}}, edited by
  \bibinfo{editor}{\bibfnamefont{S.}~\bibnamefont{Reynaud}},
  \bibinfo{editor}{\bibfnamefont{E.}~\bibnamefont{Giacobino}},
  \bibnamefont{and}
  \bibinfo{editor}{\bibfnamefont{J.}~\bibnamefont{{Zinn-Justin}}}
  (\bibinfo{publisher}{Elsevier}, \bibinfo{year}{1997}), p.
  \bibinfo{pages}{351}.

\bibitem[{\citenamefont{Goldstein et~al.}(2001)\citenamefont{Goldstein, Poole,
  and Safko}}]{goldstein_classical_2001}
\bibinfo{author}{\bibfnamefont{H.}~\bibnamefont{Goldstein}},
  \bibinfo{author}{\bibfnamefont{C.~P.} \bibnamefont{Poole}}, \bibnamefont{and}
  \bibinfo{author}{\bibfnamefont{J.~L.} \bibnamefont{Safko}},
  \emph{\bibinfo{title}{Classical Mechanics}} (\bibinfo{publisher}{Addison
  Wesley}, \bibinfo{year}{2001}), \bibinfo{edition}{3rd} ed., Chap.\ 6.

\bibitem[{\citenamefont{Koch et~al.}(2007)\citenamefont{Koch, Yu, Gambetta,
  Houck, Schuster, Majer, Blais, Devoret, Girvin, and
  Schoelkopf}}]{koch_charge-insensitive_2007}
\bibinfo{author}{\bibfnamefont{J.}~\bibnamefont{Koch}},
  \bibinfo{author}{\bibfnamefont{T.~M.} \bibnamefont{Yu}},
  \bibinfo{author}{\bibfnamefont{J.}~\bibnamefont{Gambetta}},
  \bibinfo{author}{\bibfnamefont{A.~A.} \bibnamefont{Houck}},
  \bibinfo{author}{\bibfnamefont{D.~I.} \bibnamefont{Schuster}},
  \bibinfo{author}{\bibfnamefont{J.}~\bibnamefont{Majer}},
  \bibinfo{author}{\bibfnamefont{A.}~\bibnamefont{Blais}},
  \bibinfo{author}{\bibfnamefont{M.~H.} \bibnamefont{Devoret}},
  \bibinfo{author}{\bibfnamefont{S.~M.} \bibnamefont{Girvin}},
  \bibnamefont{and} \bibinfo{author}{\bibfnamefont{R.~J.}
  \bibnamefont{Schoelkopf}}, \bibinfo{journal}{Phys. Rev. A}
  \textbf{\bibinfo{volume}{76}}, \bibinfo{pages}{042319}
  (\bibinfo{year}{2007}).

\bibitem[{\citenamefont{Schreier et~al.}(2008)\citenamefont{Schreier, Houck,
  Koch, Schuster, Johnson, Chow, Gambetta, Majer, Frunzio, Devoret
  et~al.}}]{schreier_suppressing_2008}
\bibinfo{author}{\bibfnamefont{J.~A.} \bibnamefont{Schreier}},
  \bibinfo{author}{\bibfnamefont{A.~A.} \bibnamefont{Houck}},
  \bibinfo{author}{\bibfnamefont{J.}~\bibnamefont{Koch}},
  \bibinfo{author}{\bibfnamefont{D.~I.} \bibnamefont{Schuster}},
  \bibinfo{author}{\bibfnamefont{B.~R.} \bibnamefont{Johnson}},
  \bibinfo{author}{\bibfnamefont{J.~M.} \bibnamefont{Chow}},
  \bibinfo{author}{\bibfnamefont{J.~M.} \bibnamefont{Gambetta}},
  \bibinfo{author}{\bibfnamefont{J.}~\bibnamefont{Majer}},
  \bibinfo{author}{\bibfnamefont{L.}~\bibnamefont{Frunzio}},
  \bibinfo{author}{\bibfnamefont{M.~H.} \bibnamefont{Devoret}},
  \bibinfo{author}{\bibfnamefont{S.~M.} \bibnamefont{Girvin}},
  \bibnamefont{and} \bibinfo{author}{\bibfnamefont{R.~J.}
  \bibnamefont{Schoelkopf}}, 
  \bibinfo{journal}{Phys. Rev. B}
  \textbf{\bibinfo{volume}{77}}, \bibinfo{pages}{180502}
  (\bibinfo{year}{2008}).

\bibitem[{\citenamefont{Bouchiat et~al.}(1998)\citenamefont{Bouchiat, Vion,
  Joyez, Esteve, and Devoret}}]{bouchiat_quantum_1998}
\bibinfo{author}{\bibfnamefont{V.}~\bibnamefont{Bouchiat}},
  \bibinfo{author}{\bibfnamefont{D.}~\bibnamefont{Vion}},
  \bibinfo{author}{\bibfnamefont{P.}~\bibnamefont{Joyez}},
  \bibinfo{author}{\bibfnamefont{D.}~\bibnamefont{Esteve}}, \bibnamefont{and}
  \bibinfo{author}{\bibfnamefont{M.~H.} \bibnamefont{Devoret}},
  \bibinfo{journal}{Phys. Scr.} \textbf{\bibinfo{volume}{T76}},
  \bibinfo{pages}{165} (\bibinfo{year}{1998}).

\bibitem[{\citenamefont{Nakamura et~al.}(1999)\citenamefont{Nakamura, Pashkin,
  and Tsai}}]{nakamura_coherent_1999}
\bibinfo{author}{\bibfnamefont{Y.}~\bibnamefont{Nakamura}},
  \bibinfo{author}{\bibfnamefont{Y.~A.} \bibnamefont{Pashkin}},
  \bibnamefont{and} \bibinfo{author}{\bibfnamefont{J.~S.} \bibnamefont{Tsai}},
  \bibinfo{journal}{Nature {(London)}} \textbf{\bibinfo{volume}{398}},
  \bibinfo{pages}{786} (\bibinfo{year}{1999}).

\bibitem[{\citenamefont{Vion et~al.}(2002)\citenamefont{Vion, Aassime, Cottet,
  Joyez, Pothier, {C.Urbina}, Esteve, and
  Devoret}}]{vion_manipulatingquantum_2002}
\bibinfo{author}{\bibfnamefont{D.}~\bibnamefont{Vion}},
  \bibinfo{author}{\bibfnamefont{A.}~\bibnamefont{Aassime}},
  \bibinfo{author}{\bibfnamefont{A.}~\bibnamefont{Cottet}},
  \bibinfo{author}{\bibfnamefont{P.}~\bibnamefont{Joyez}},
  \bibinfo{author}{\bibfnamefont{H.}~\bibnamefont{Pothier}},
  \bibinfo{author}{\bibnamefont{{C.Urbina}}},
  \bibinfo{author}{\bibfnamefont{D.}~\bibnamefont{Esteve}}, \bibnamefont{and}
  \bibinfo{author}{\bibfnamefont{M.~H.} \bibnamefont{Devoret}},
  \bibinfo{journal}{Science} \textbf{\bibinfo{volume}{296}},
  \bibinfo{pages}{886} (\bibinfo{year}{2002}).

\bibitem[{\citenamefont{Metcalfe et~al.}(2007)\citenamefont{Metcalfe, Boaknin,
  Manucharyan, Vijay, Siddiqi, Rigetti, Frunzio, Schoelkopf, and
  Devoret}}]{metcalfe_measuringdecoherence_2007}
\bibinfo{author}{\bibfnamefont{M.}~\bibnamefont{Metcalfe}},
  \bibinfo{author}{\bibfnamefont{E.}~\bibnamefont{Boaknin}},
  \bibinfo{author}{\bibfnamefont{V.}~\bibnamefont{Manucharyan}},
  \bibinfo{author}{\bibfnamefont{R.}~\bibnamefont{Vijay}},
  \bibinfo{author}{\bibfnamefont{I.}~\bibnamefont{Siddiqi}},
  \bibinfo{author}{\bibfnamefont{C.}~\bibnamefont{Rigetti}},
  \bibinfo{author}{\bibfnamefont{L.}~\bibnamefont{Frunzio}},
  \bibinfo{author}{\bibfnamefont{R.~J.} \bibnamefont{Schoelkopf}},
  \bibnamefont{and} \bibinfo{author}{\bibfnamefont{M.~H.}
  \bibnamefont{Devoret}}, \bibinfo{journal}{Phys. Rev. B}
  \textbf{\bibinfo{volume}{76}}, \bibinfo{pages}{174516}
  (\bibinfo{year}{2007}).

\bibitem[{\citenamefont{Zorin et~al.}(1996)\citenamefont{Zorin, Ahlers,
  Niemeyer, Weimann, and Wolf}}]{zorin_background_1996}
\bibinfo{author}{\bibfnamefont{A.~B.} \bibnamefont{Zorin}},
  \bibinfo{author}{\bibfnamefont{F.}~\bibnamefont{Ahlers}},
  \bibinfo{author}{\bibfnamefont{J.}~\bibnamefont{Niemeyer}},
  \bibinfo{author}{\bibfnamefont{T.}~\bibnamefont{Weimann}}, 
  \bibinfo{author}{\bibfnamefont{H.}~\bibnamefont{Wolf}},
  \bibinfo{author}{\bibfnamefont{V.~A.}~\bibnamefont{Krupenin}},  
  \bibnamefont{and}
  \bibinfo{author}{\bibfnamefont{S.~V.}~\bibnamefont{Lothkov}},
  \bibinfo{journal}{Phys. Rev. B} \textbf{\bibinfo{volume}{53}},
  \bibinfo{pages}{13682} (\bibinfo{year}{1996}).
  
\bibitem[{\citenamefont{Kafanov et~al.}(2008)\citenamefont{Kafanov et al.}}]{kafanov}
\bibinfo{author}{\bibfnamefont{S.} \bibnamefont{Kafanov}},
  \bibinfo{author}{\bibfnamefont{H.}~\bibnamefont{Brenning}},
  \bibinfo{author}{\bibfnamefont{T.}~\bibnamefont{Duty}},
  \bibnamefont{and}
  \bibinfo{author}{\bibfnamefont{P.}~\bibnamefont{Delsing}},
  \bibinfo{journal}{Phys. Rev. B} \textbf{\bibinfo{volume}{78}},
  \bibinfo{pages}{125411} (\bibinfo{year}{2008}).
    

\bibitem[{\citenamefont{White et~al.}(2009)\citenamefont{White, Pasienski,
  {McKay}, Zhou, Ceperley, and {DeMarco}}}]{white_strongly_2009}
\bibinfo{author}{\bibfnamefont{M.}~\bibnamefont{White}},
  \bibinfo{author}{\bibfnamefont{M.}~\bibnamefont{Pasienski}},
  \bibinfo{author}{\bibfnamefont{D.}~\bibnamefont{{McKay}}},
  \bibinfo{author}{\bibfnamefont{S.~Q.} \bibnamefont{Zhou}},
  \bibinfo{author}{\bibfnamefont{D.}~\bibnamefont{Ceperley}}, \bibnamefont{and}
  \bibinfo{author}{\bibfnamefont{B.}~\bibnamefont{{DeMarco}}},
  \bibinfo{journal}{Phys. Rev. Lett.} \textbf{\bibinfo{volume}{102}},
  \bibinfo{pages}{055301} (\bibinfo{year}{2009}).
  
  
\bibitem[{\citenamefont{Lee}(1994)\citenamefont{Lee and Chalker}}]{lee94}
\bibinfo{author}{\bibfnamefont{D.~K.~K.}~\bibnamefont{Lee}}, \bibnamefont{and}
  \bibinfo{author}{\bibfnamefont{J.~T.}~\bibnamefont{{Chalker}}},
  \bibinfo{journal}{Phys. Rev. Lett.} \textbf{\bibinfo{volume}{72}},
  \bibinfo{pages}{1510} (\bibinfo{year}{1994}).
  
\bibitem[{\citenamefont{Aronov}(1994)\citenamefont{Aronov et al.}}]{aronov94}
\bibinfo{author}{\bibfnamefont{A.~G.}~\bibnamefont{Aronov}},
\bibinfo{author}{\bibfnamefont{A.~D.}~\bibnamefont{Mirlin}}, \bibnamefont{and}
\bibinfo{author}{\bibfnamefont{P.}~\bibnamefont{W\"olfle}}, 
  \bibinfo{journal}{Phys. Rev. B} \textbf{\bibinfo{volume}{49}},
  \bibinfo{pages}{16609} (\bibinfo{year}{1994}).  

\bibitem[{\citenamefont{Merzbacher}(1997)}]{merzbacher_quantum_1997}
\bibinfo{author}{\bibfnamefont{E.}~\bibnamefont{Merzbacher}},
  \emph{\bibinfo{title}{Quantum Mechanics}} (\bibinfo{publisher}{Wiley},
  \bibinfo{year}{1997}), \bibinfo{edition}{3rd} ed.

\bibitem[{\citenamefont{Weinberg}(2005)}]{weinberg_quantum_2005}
\bibinfo{author}{\bibfnamefont{S.}~\bibnamefont{Weinberg}},
  \emph{\bibinfo{title}{The Quantum Theory of Fields, Volume 1: Foundations}}
  (\bibinfo{publisher}{Cambridge University Press}, \bibinfo{year}{2005}), Chap.\ 1 and Appendix A.

\bibitem[{\citenamefont{Wigner}(1931)}]{wigner_gruppentheorie_1931}
\bibinfo{author}{\bibfnamefont{E.~P.} \bibnamefont{Wigner}},
  \emph{\bibinfo{title}{Gruppentheorie und ihre Anwendung auf die
  Quantenmechanik der Atomspektren}} (\bibinfo{publisher}{Vieweg},
  \bibinfo{address}{Braunschweig}, \bibinfo{year}{1931}).

\bibitem[{\citenamefont{Rogge and Haug}(2009)}]{rogge_three_2009}
\bibinfo{author}{\bibfnamefont{M.~C.} \bibnamefont{Rogge}} \bibnamefont{and}
  \bibinfo{author}{\bibfnamefont{R.~J.} \bibnamefont{Haug}},
  \bibinfo{journal}{New J. Phys.} \textbf{\bibinfo{volume}{11}},
  \bibinfo{pages}{113037} (\bibinfo{year}{2009}).

\end{thebibliography}
\end{document}